\documentclass[final]{IEEEtran}
\usepackage{hyperref}

\usepackage{xcolor}
\usepackage{xspace}
\usepackage[sort,compress]{cite}
\usepackage{import}     % Nested file importing 
\usepackage{pgfplots}   % high-quality plots
\pgfplotsset{compat=1.14} % use version 1.15 of pgfplots (resolves a latex warning)
\usepackage{psfrag}     % place text on figures
\usepackage[autostyle, english = american]{csquotes}
\usepackage[framemethod=tikz,xcolor=true]{mdframed} % for drawing boxes around the table and figures
\usepackage{makecell}   % support for inserting multi-line text
\usepackage{amsfonts,amsmath,amssymb}   % provides math typesetting
\usepackage{subcaption}
\usepackage[font={small,up}]{caption}
\usetikzlibrary{patterns}
\usepackage[ruled]{algorithm2e}
\usepackage{dirtytalk} % macros for quotations (\say{})

\graphicspath{{figures/}{figures/source/}}

\newcommand{\secref}[1]{\mbox{Section~\ref{#1}}}
\newcommand{\figref}[1]{\mbox{Figure~\ref{#1}}}
\newcommand{\tabref}[1]{\mbox{Table~\ref{#1}}}
\newcommand{\eqnref}[1]{\mbox{Equation~\ref{#1}}}
\newcommand{\algref}[1]{\mbox{Algorithm~\ref{#1}}}

\newcommand{\appref}[1]{\mbox{Appendix~\ref{#1}}}

\MakeOuterQuote{"} % ensures quotation marks formatted correctly

\newcommand{\etal}{et~al.\@\xspace}
\newcommand{\eg}{e.g.,\@\xspace}
\newcommand{\ie}{i.e.,\@\xspace}

\usepackage{booktabs} % quality table styles
\usepackage{multirow} %%multi-row
\newcommand{\ra}[1]{\renewcommand{\arraystretch}{#1}} % row-adjust: manually adjust space between table rows

\DeclareMathOperator*{\argmin}{arg\,min}
\usepackage{tikz}
\newcommand*\circled[1]{\tikz[baseline=(char.base)]{
            \node[shape=circle,draw,inner sep=1pt] (char) {#1};}}
            
\makeatletter
\def\namedlabel#1#2{\begingroup
    #2%
    \def\@currentlabel{#2}%
    \phantomsection\label{#1}\endgroup
}
\makeatother

\begin{document}

\title{Privacy Partitioning: Protecting User Data During the Deep Learning Inference Phase}

\author{\IEEEauthorblockN{Jianfeng Chi\IEEEauthorrefmark{1},
Emmanuel Owusu\IEEEauthorrefmark{2}, Xuwang Yin\IEEEauthorrefmark{1}, Tong Yu\IEEEauthorrefmark{2}, William Chan\IEEEauthorrefmark{3}, Patrick Tague\IEEEauthorrefmark{2}, and Yuan Tian\IEEEauthorrefmark{1}}
\IEEEauthorblockA{\{jc6ub, xy4cm, yuant\}@virginia.edu,
\{eowusu, tague\}@cmu.edu, tong.yu@sv.cmu.edu, wchan212@gmail.com\\
\IEEEauthorrefmark{1}University of Virginia, 
\IEEEauthorrefmark{2}Carnegie Mellon University, 
\IEEEauthorrefmark{3}Google Brain\\
}}

% \and
% \IEEEauthorblockN{James Kirk\\ and Montgomery Scott}
% \IEEEauthorblockA{Starfleet Academy\\
% someemail@somedomain.com}


%Telephone: (800) 555--1212, Fax: (888) 555--1212}


\maketitle

\begin{abstract}
    \boldmath %Abstract
We present a practical method for protecting data during the inference phase of deep learning based on bipartite topology threat modeling and an interactive adversarial deep network construction. We term this approach \emph{Privacy Partitioning}. 
In the proposed framework, we split the machine learning models and deploy a few layers into users' local devices, and the rest of the layers into a remote server. We propose an approach to protect user's data during the inference phase, while still achieve good classification accuracy. 

We conduct an experimental evaluation of this approach on benchmark datasets of three computer vision tasks. The experimental results indicate that this approach can be used to significantly attenuate the capacity for an adversary with access to the state-of-the-art deep network's intermediate states to learn privacy-sensitive inputs to the network. For example, we demonstrate that our approach can prevent attackers from inferring the private attributes such as gender from the Face image dataset without sacrificing the classification accuracy of the original machine learning task such as Face Identification.    
\end{abstract}

\section{Introduction}\label{partition:sec:introduction}
% Introduction

Presently, deep learning has been established as the most widely used machine learning solution. The popularity of the deep  neural network (DNN) is due in large part to its effectiveness and applicability to a wide variety of complex tasks, and its relative ease of use compared to other machine learning solutions~\cite{chung2017lip,suwajanakorn2017synthesizing,wojna2017attention, santoro2017simple,isola2017image}. The efficacy of the deep learning solutions have also benefited significantly from broad technology trends such as improved computational power, increased inter-networking capacity, and a proliferation of massive structured datasets and data streaming sources.

The capacity for deep networks to leverage massive datasets has drawn interest from a privacy-leery public and security researchers -- leading to proposals for privacy-preserving protection mechanisms. However, recent works on protecting data privacy of deep learning~\cite{mcmahan2016communication,shokri2015privacy} largely focus on protecting training data during the model learning phase using differential privacy, whereas deep learning services potentially pose a more significant privacy threat during the model inference phase where more user data is processed. Recently, cryptography-based protocols are also proposed~\cite{liu2017oblivious, gilad2016cryptonets} to protect data privacy in model inference phase. However, these protocols impose computation overhead on model inference and it is not feasible for those deep models with cryptography-based protocols to be deployed in computational low-performing devices such as IoT devices. Researchers also proposed differential-privacy-based solutions for protecting aggregated data during inference phase. However, these solutions don't protect individual user data during the model inference phase. 

In this work, we propose a new framework for protecting user data during the model inference phase where users use their data to get classification results. The output of a deep learning neural network construction is comprised of multiple intermediate layers that encode information regarding the previous layers, providing a channel for unauthorized access to privacy-sensitive data. We consider the case that an adversary whose goal is to recover input data or extract sensitive information from the input data for the request of deep learning services and the adversary can have complete access to the remote party who offers the deep learning services. For example, a malware on the cloud server can get access to the intermediate layers of a face detection classifier to infer the emotions of the users when the user just wants the camera in their smartphone to track their face. 

\begin{figure}[t!]
\begin{mdframed}[
    tikzsetting={align=center, draw=black, thick, align=center},
    innerrightmargin=5pt,innerleftmargin=5pt,innerbottommargin=5pt, topline=false,
  rightline=false,
  leftline=false,
  bottomline=false
]
    
    \centering
    
    \psfrag{a}[c][c][0.9]{$\Theta_{l}$}
    \psfrag{b}[c][c][0.9]{$\Theta_{r}$}
    
    \psfrag{c}[c][c][0.9]{$x_0$}
    \psfrag{d}[c][c][0.9]{$x_1$}
    \psfrag{e}[c][c][0.9]{\vdots}
    \psfrag{f}[c][c][0.9]{$x_n$}
    
    \psfrag{g}[c][c][0.9]{$y_0$}
    \psfrag{h}[c][c][0.9]{\vdots}
    \psfrag{i}[c][c][0.9]{$y_n$}
    
    \psfrag{j}[c][c][0.9]{$\mathcal{L}_{i}$}
    \psfrag{k}[c][c][0.9]{$\mathcal{L}_{i+1}$}
    
    \psfrag{l}[c][c][0.9]{$h_0$}
    \psfrag{m}[c][c][0.9]{$h_1$}
    \psfrag{n}[c][c][0.9]{\vdots}
    \psfrag{o}[c][c][0.9]{$h_n$}

    \includegraphics[width=3.33in]{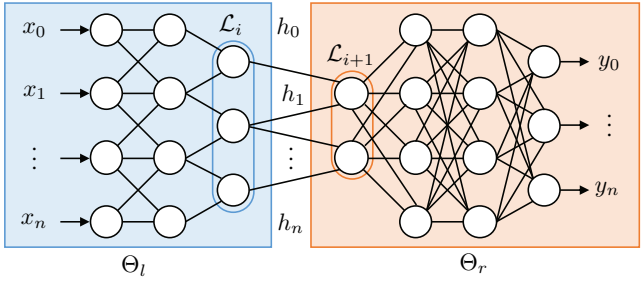}
    \caption{ {\bf Bipartite Deep Network Topology:} This figure depicts a bipartite deployment context and privacy model applied to deep convolutional neural network $\Theta$. Partition $\Theta_{l}$ is the \emph{local} computing context. Partition $\Theta_{r}$ is the \emph{remote} computing context. Hidden state $\mathcal{H}$ is the output of the last local transformation layer $\mathcal{L}_{i}$ and the input of the first remote transformation layer $\mathcal{L}_{i+1}$. Given deep network $\Theta$ and activation state $\mathcal{H}$, this framework generates bipartite network $\{\Theta_{l},\Theta_{r}\}$ that learns model $f_{\theta}:\mathcal{X}\rightarrow\mathcal{Y}$ while attenuating the capacity of attackers to learn function $f_{\theta_a}:\mathcal{H}\rightarrow\mathcal{X}$ and private attributes from the recovered inputs.}
    
    \label{partition:fig:topology}
    
\end{mdframed}
\end{figure}

We then present a practical method for attenuating the privacy risk potential of intermediate layers in deep learning inference tasks -- a framework we term \emph{Privacy Partitioning}. Deep neural networks $\Theta$ in this framework are partitioned resulting in bipartite network \{$\Theta_l, \Theta_r$\} and access to the network's inputs it restricted to $\Theta_l$ (\figref{partition:fig:topology}). During the model training phase, we optimize the machine learning model to only share information about the expected utility of the data such as face identification, and avoid disclosing private information such as the emotional information to the remote layers. As a result, when the users are using the model during the inference phase, their emotional data will not be easily extracted from the remote layers. 

Privacy Partitioning attenuates the privacy risk potential of one part of a deep network (\ie remote partition $\Theta_{r}$) to the other part of the deep network (\ie local partition $\Theta_{l}$). The privacy partitioning framework is itself an instance of centralized learning with key adaptations that support a bipartite data privacy threat model. This adaptation compatible and composable with other deep learning security and privacy protection mechanisms~\cite{mcmahan2016communication,shokri2015privacy}, enables the service provider to protect trade secrets (\eg the intellectual property of DNNs) more readily, and enables machine learning engineers to readily implement privacy partitions without the additional learning curve of cryptographic protection.

The privacy partitioning framework is composed of two primary components: (i) a bipartite topology (\figref{partition:fig:topology}) and (ii) an interactive adversarial deep neural network construction (\figref{partition:fig:model}). These two aspects enable a two-part data privacy and deployment model where the local partition $\Theta_{l}$ is restricted to authorized users whereas the remote partition $\Theta_{r}$ may be safely entrusted to honest-but-curious remote parties and deployed in a potentially malicious computing context. 

The bipartite topology is motivated by it's apt applicability to many computing use cases featuring a local-vs-remote bipartite threat model including client to server context, local edge network to remote cloud contexts, and use cases requiring offloading from constrained devices to untrusted resource-rich proximate infrastructure. The interactive adversarial approach to constructing a privacy partitioning (the focus of \secref{partition:sec:framework}) is inspired in part by cybernetics theory -- where learning a skill is understood to be a feedback and refinement process -- and by game theory -- where optimal decision-making is arrived at through iterative simulations of independent and competing actors in a strategic game. In our context of privacy partitionings, the skill the deep network is tasked with learning is to adapt a reference network in such a way that the remote partition receives less information about inputs while maintaining the accuracy of the reference network. Similarly, the strategic game is between a defense oracle $\Theta_{d}$, competing to battle harden bipartite network \{$\Theta_{l}, \Theta_{r}$\} by optimizing for recovery accuracy before the convergence of the Stochastic Gradient Descent (SGD) algorithm.
 
\noindent \textbf{Contributions.} In this study we present Privacy Partitioning, a practical privacy protection framework for model inference phase. In summary, this work makes the following contributions:

\begin{itemize}
    
    \item We propose a novel framework for learning accurate deep networks that are resilient against input recovery attacks during model inference phase (\secref{partition:sec:framework}). 
        
    \item Unlike past research for protecting aggregated data during model inference phase, we propose a solution to achieve a balance between the utility and privacy of the individual data for model inference (\secref{partition:sec:threat-model}).
    
    \item We experimentally demonstrate the effectiveness of our approach. In all experiments, the results indicate that the privacy partitioning framework can significantly reduce the privacy risk potential of using deep network activation states to learn the inputs to the network (\secref{partition:sec:experiments}).
    
\end{itemize}

\noindent \emph{\bf Organization.} The topic of the next section, \secref{partition:sec:threat-model}, is the problem definition including the deployment model, adversary model, assumptions, and desired properties. \secref{partition:sec:bg} covers some preliminary background information regarding our work. \secref{partition:sec:framework} presents the proposed framework. \secref{partition:sec:experiments} presents experimental results for the privacy protection mechanism proposed here using three computers vision classifier networks. \secref{partition:sec:related-work} discusses the related literature and \secref{partition:sec:conclusion} presents a summary of conclusions.

\section{Problem Definition}\label{partition:sec:threat-model}
%---------------------------------
\subsection{Model and Assumptions}
%---------------------------------

\noindent \emph{\bf Deployment Model.}

Our use case defines outsourced computation in the sense advocated by the hybrid public cloud and edge computing deployment models. In particular, we consider the case where the users want to get classification results of their data from a remote server and still want to protect the privacy of their data. Accordingly, we identify the following four key stakeholders and their primary roles:

(i) The \textbf{\emph{ cloud provider}} (remote admin) offers data center-based compute, storage, and network resources (\ie the remote computing context) as-a-service to cloud customers who wish to lease them for a certain amount of time.

(ii) The \textbf{\emph{ edge provider}} (local admin) offers edge network-based compute, storage, and network resources (\ie the local computing context) as-a-platform for use by the network of co-located users and embedded devices. Local admins are interested in ensuring that all services deployed within the local domain adhere to its security and privacy requirements.

(iii) The \textbf{\emph{ service provider}} (app) wishes to lease both edge computing and cloud computing resources to deploy services with protections that are robust against the threats indicated in the adversary model.

Finally, (iv) the \textbf{\emph{end user}} (user device) wishes to use authorized apps, guaranteeing the integrity and confidentiality of network interactions within the local domain and the privacy of data requested by shared deep learning models.

\noindent \emph{\bf Adversary Model.}

In the case of centralized learning, the entirety of deep network $\Theta$ is deployed to a single computing context managed by a single administrative domain. The threat model defined here involves a bipartite topology and privacy policy managed by two administrative domains: the data custodian (local admin) and the cloud computing provider (remote admin). This threat model addresses a commonly occurring deployment context and data privacy requirement where a data provenance/handling domain would like to safely leverage remote compute resources on behalf of its users. 

The local admin, would like to restrict data access to the local computing context. The remote admin, provides the computing resources comprising the remote computing context. Bipartite learning results in transformation of $\Theta$ to \{$\Theta_{l}\cdot\Theta_{r}$\} where the local partition $\Theta_{l}$ is the portion of deep network $\Theta$ managed by the local admin and the remote partition $\Theta_{r}$ is the portion of deep network $\Theta$ managed by the remote admin (see \figref{partition:fig:topology}).    

We consider the privacy of inputs to deep network \{$\Theta_{l}\cdot\Theta_{r}$\} during the inference phase given an adversary with complete access to the remote computing context $\Theta_{r}$. The goal of the adversary is to recover input data that contains private attributes given outputs of the local partition $\Theta_{l}$ (\ie intermediate state), access to all hidden states generated within $\Theta_{r}$, and access to network output. Thus the adversary could be an attacker who gains access to the traffic between the $\Theta_{l}$ and $\Theta_{r}$, a malicious agent who gains unauthorized access to the remote computing context, or a remote admin who wants to infer more information from user data.

\noindent \emph{\bf Assumptions.}

A modicum level of trust for the servers managing the process is required. The servers comprising the local computing context are trusted to securely collect user inputs as well as to securely manage the cryptographic assets and access control policies of the individuals and devices within the local domain. Thus we consider data compromises resulting from the local computing context out of scope. Additionally, we consider side-channel style inference attacks beyond the scope of this work. The servers constituting local and remote computing contexts are trusted to operate normally (\ie we consider the denial of service style attacks out of scope).

We assume that bipartite network \{$\Theta_{l}\cdot\Theta_{r}$\} is trained using the process described in \secref{partition:sec:framework} and experimentally evaluated in \secref{partition:sec:experiments}. This threat model deals with data privacy protection during the model inference phase. Therefore, the attacker may know, may partially know, or may not know the dataset that is used during the training phase to train and validate bipartite network \{$\Theta_{l}\cdot\Theta_{r}$\}. Although this threat model does not assume that the attacker has access to training data, we evaluate the privacy partitioning framework in \secref{partition:sec:experiments} using the strong attackers who does utilize the same training data that was used to learn \{$\Theta_{l}\cdot\Theta_{r}$\}.

\subsection{Desired Properties}\label{partition:sec:des_properties}
\noindent The following list contains the desired properties for the Privacy Partitioning.

\begin{enumerate}

\item[\namedlabel{data:prop:utility}{$P 1$}] {\bf Utility.}  We would like the following objectives to be satisfied:
    
    \begin{enumerate}
    
    \item[\namedlabel{data:prop:performant}{$P 1.1$}] \emph{Performant}. Protections result in a negligible reduction to model accuracy.
    
    \item[\namedlabel{data:prop:learning-based}{$P 1.2$}] \emph{Learning-Based}. Protections do not require processes that are outside of the deep learning tool-set such as managing cryptographic secrets.
    
    \end{enumerate}

\item[\namedlabel{data:prop:versatility}{$P 2$}] {\bf Versatility.}  We would like the following objectives to be satisfied:

    \begin{enumerate}

    \item[\namedlabel{data:prop:compatible}{$P 2.1$}] \emph{Compatible}. Protections can be readily applied to existing deep networks.
    
    \item[\namedlabel{data:prop:complementary}{$P 2.2$}] \emph{Complementary}. Protections can be readily deployed alongside end-to-end security and privacy protection mechanisms.
    
    \end{enumerate}

\end{enumerate}

\section{Background}\label{partition:sec:bg}
%Background

In this section, we present several background topics related to the basics of deep learning and the deep learning deployment typologies in collaborative setting.

\subsection{The Basics of Deep Learning}
In our work, we focus on the setting of supervised learning for simplicity. A deep learning model $f_{\theta}: \mathcal{X} \rightarrow \mathcal{Y}$ parameterized by $\theta$. For the classification problems, $\mathcal{X}$ is high dimensional vector space and $\mathcal{Y}$ is the space for the classes. Given a labeled dataset $\{ (x_i, y_i) \}_{i=1}^{m^\prime}$ where $(x_i, y_i) \in \mathcal{X} \times \mathcal{Y}$. The dataset is usually partitioned into training data of size $m$ and test data.

In order to learn a good DNN model that perform well on the test data, we will try to minimize the loss function $l$ which measures the difference between ground truth labels and the predicted labels:

\begin{equation}
	\nonumber
    \min_{\theta} \frac{1}{m} \sum_{i=1}^{m} l\big(y_i, f_{\theta}(x_i)\big)
\end{equation}

The loss function is usually optimized gradient-based optimization algorithm such as SGD. Popular choices of $f_{\theta}$ in the application of deep learning include Multilayer Perceptrons (MLPs), Convolutional Neural Networks (CNNs) and Recurrent Neural Networks (RNNs).

\subsection{Deep Learning in Collaborative Setting}

Consider the case where $N$ users of a mobile app each collect and store personal data locally on their mobile devices.
The goal of collaborative deep learning is to leverage a deep network that incorporates the data of all $N$ user devices to, for example, produce relevant personalized services for each user. 

Due to the potentially sensitive nature of personal data we would like to achieve a high-performance deep network while simultaneously providing strong data privacy guarantees for all participants. That is, we would like to maximize the accuracy of learned model while minimizing the amount of individual attributes that can be leaked or inferred about participant- we would like to establish \emph{privacy-preserving} collaborative learning. Individual attributes include recovered raw input data as well associated personally-identifiable user attributes such as identity, interests, habits, and social network.

In general, there are two architectural approaches to learn large-scale deep networks: \textit{centralized learning} and \textit{distributed learning}. The choice between the two (as well as the spectrum of topologies spanning them) have differing implications for performance, scalability, ease of deployment, and data privacy.

\noindent \emph{\bf Centralized Learning.} Traditionally, machine learning frameworks have assumed access to a centralized repository of data (or otherwise considered data mining and curation tasks as independent of, and generally beyond the scope of, model learning) even in the case of collaborative learning tasks requiring access to massive personally-identifiable privacy-sensitive datasets. 

The centralized approach is developer-friendly, since a single operator can manage the entire process. Production-grade deep networks will often utilize protocol-agnostic data protection measures such as end-to-end encryption for secure data transfer over public networks and organizational consumer protection measures such as terms of service agreements, enabling the deep network itself to be free of complexity and performance overheads due to protocol-level support for privacy protection. However, these peripheral protection measures provide data confidentiality instead of data secrecy. Thus, even in the best case, these measures do not directly diminish the potential for privacy intrusion or large scale mishandling of personal information--instead relying on implicit trust agreements between individuals and service providers and on accountability measures for deterrence and recompense. In the worst case, large central repositories of highly-structured data provide significant points of failure for unauthorized access.

\noindent \emph{\bf Distributed Learning.} Researchers have proposed a variety of decentralized architectures as solutions to both performance and privacy. Decentralized deep learning (also known as distributed or federated deep learning) includes proposals for building and updating a unified model without the need to store individual data in the cloud and proposals that combine both a personalized individual model managed locally on user devices and a shared model constructed from anonymity averaged data~\cite{mcmahan2016communication,shokri2015privacy}.

In principle, distributed learning approaches enable participants to enjoy the full benefits of rich shared models without the need to centrally stored data. This benefit often comes at the expense of increased complexity due to asynchronous operations and an assorted computing and data management context. Additionally, fully distributed learning implementations may complicate attempts to protect the trade secrets and intellectual property of the DNNs.

\section{Framework}\label{partition:sec:framework}
%Proposed Framework

In this section, we present the privacy partitioning framework: a novel method for learning accurate deep networks that are resilient against input recovery attacks.

\subsection{Bipartite Topology Design}

\begin{figure*}[ht]
\begin{mdframed}[
    tikzsetting={align=center, draw=black, thick, align=center},
    innerrightmargin=5pt,innerleftmargin=5pt,innerbottommargin=5pt, topline=false,
  rightline=false,
  leftline=false,
  bottomline=false
]    
    \centering
    %\psfrag{a}[l][c][0.9]{$\mathcal{X} = x_{i=1, \dots, m}$}
    \psfrag{a}[c][c][0.9]{$\mathnormal{X}$}
    \psfrag{b}[c][c][0.9]{$\Theta_{l}$}
    \psfrag{c}[c][c][0.9]{$\mathnormal{H}$}
    \psfrag{d}[c][c][0.9]{$\Theta_{r}$}
    \psfrag{e}[c][c][0.9]{$\mathnormal{Y}$}
    \psfrag{f}[c][c][0.9]{$\Theta_{d}$}
    \psfrag{g}[c][c][0.9]{$\mathnormal{X}^\prime$}
    
    \psfrag{h}[c][c][0.9]{(a)}
    
    \psfrag{i}[c][c][0.9]{$\mathnormal{X}$}
    \psfrag{j}[c][c][0.9]{$\Theta_{l}$}
    \psfrag{k}[c][c][0.9]{$\mathnormal{H}$}
    \psfrag{l}[c][c][0.9]{$\Theta_{r}$}
    \psfrag{m}[c][c][0.9]{$\mathnormal{Y}$}
    \psfrag{n}[c][c][0.9]{$\Theta_{a_0}$}
    \psfrag{o}[c][c][0.9]{$\Theta_{a_1}$}
    \psfrag{p}[c][c][0.9]{$\dotsc$}
    \psfrag{q}[c][c][0.9]{$\Theta_{a_k}$}
    \psfrag{r}[c][c][0.9]{$\mathnormal{X}^\prime_0$}
    \psfrag{s}[c][c][0.9]{$\mathnormal{X}^\prime_1$}
    \psfrag{t}[c][c][0.9]{$\dotsc$}
    \psfrag{u}[c][c][0.9]{$\mathnormal{X}^\prime_n$}
    
    \psfrag{v}[c][c][0.9]{(b)}
    
    \psfrag{x}[l][c][0.7]{\texttt{Learning Phase}}
    \psfrag{y}[l][c][0.7]{\texttt{Inference Phase}}

    \includegraphics[width=7in]{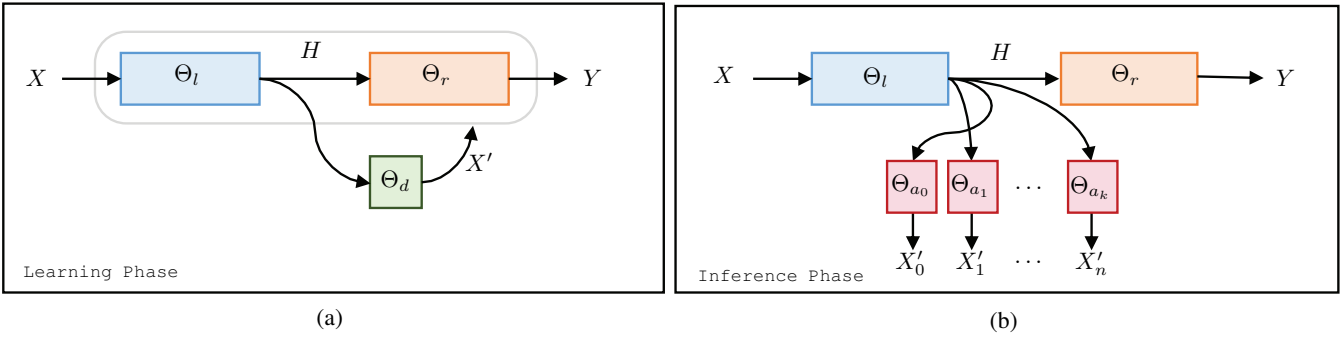}
    \caption{{\bf Model Learning and Model Inference:} This figure illustrates the two phases of deploying privacy partition protections to a deep network: (a) the learning phase  and (b) the inference phase. The learning phase encompasses several machine learning activities including parameter tuning using training and validation datasets. The inference phase refers to processing inputs from end users after the network has been deployed. During the learning phase, deep network $\Theta$ is first partitioned into a bipartite topology \{$\Theta_{l}$, $\Theta_{r}$\} and defender $\Theta_{d}$, acting as an oracle for attacker behavior, is added as a feedback mechanism for minimizing the potential of recovering input $X$ via hidden state $H$ (see \eqnref{partition:eq:defender-feedback}). During the inference phase, when bipartite network \{$\Theta_{l}$, $\Theta_{r}$\} has been deployed, the adversary devises counterpart networks that each attempt to recover $X$ from $H$.} 
    \label{partition:fig:model}
    
\end{mdframed}
\end{figure*}

We consider the case where a local admin manages $N$ devices. We define a local computing context managed by the local admin in the sense advocated by the mobile edge computing and fog computing paradigms--where the local computing context represents a spatiotemporal domain of data streaming sources such as a connected car, a smart home, an enterprise network, a university campus, a smart city, or connected regional infrastructure. In the context of spatiotemporally defined computing domains, a local domain represents a single operating domain even though this domain may contain many data streaming elements all operated by different users and all serving different purposes (\eg a smart home with connected embedded devices, network elements, displays, CPUs, sensors, and actuators). 

These $N$ devices fall under the domain of the local computing context due to their proximity to the same set of users. Collectively, the devices of local computing context collect and store personal and contextual data regarding the users and activities within the domain. The local computing context is also responsible for sending out processed user's data which the user ask for machine learning services to the remote server. In the remainder of the paper, we use the terms local computing context, local domain, and local computing nodes interchangeably. We use the term local partition (denoted by $\Theta_{l}$) to refer to the part of the deep network that is run on the local computing context. Similarly, we use the the terms remote computing context, remote domain, and remote computing nodes interchangeably. And we use the term remote partition (denoted by $\Theta_{r}$) to refer to the part of the deep network that is run on the remote computing context.

Please note that, even though we present privacy partitionings in the context of local area networks comprised of co-present connected devices, privacy partitionings may be applied in any other contexts that call for bipartite privacy policy with a privacy-hardened local computing context. In general, the ideal problems utilizing the privacy partitioning framework have these aspects: (i) deep learning tasks where it is required or advantageous to decouple data handling and model inference tasks (ii) when the data entrusted to a deep network is privacy-sensitive (iii) when the deep learning model is sufficiently complex to see performance increases from outsourced computation (iv) when it's advantageous to support network-layer-based or user-domain-based controls that adjust the trade-off between privacy and performance.%\todo{need to brainstorm and carefully select this list of conditions for using our framework}

Another component of our framework is the remote domain. The remote domain receives processed user data for machine learning services and also provide the end-users with machine learning services. Due to the contextual and potentially sensitive nature of this data, the remote domain is trusted to run deep learning model $\Theta$ but not trusted with access to raw user data. The high-level goal of the framework is to train a highly-accurate deep learning model that incorporates personal and contextual data from a local domain while leveraging remote compute resources residing in the remote domain for deep learning services. Next, we describe an \textit{interactive adversarial deep network construction} designed to achieve this end.

%\subsection{Interactive Adversarial Deep Neural Network Construction}
\subsection{Model Learning Phase}

Intuitively, an initial motivation for computing a portion of the deep network locally is because the hidden states of the successive layers contain more transformations, resulting in less a traceable representation with regards to the attributes of the original inputs to the network.

However, this initial measure is not enough because an attacker can, in many cases, generate a fairly accurate estimation of the raw input data set $\mathcal{X}$ from the output of the local layers $\mathcal{H}$ (\ie the attacker can compute $f_{\theta_a}:\mathcal{H}\rightarrow\mathcal{X}$.

A privacy partitioned deep network should meet two goals. (i) Bipartite network $\{\Theta_{l},\Theta_{r}\}$ should be a functional approximation of reference deep network $\Theta$--\ie the resulting model should achieve good performance. (ii) It should be prohibitively difficult for an attacker to recover raw input data given the output of local layers. Therefore, when training the models, we need to optimize the objective function so that model performance is maintained as the potential for an input recovery is lessened.

In effect, we would like to prevent the attacker from recovering the inputs by ensuring the local layer operations are irreversible. In order to achieve this, we introduce an additional component into the model learning phase: \textit{defender} ($\Theta_{d}$). The role of the defender is to simulate the attackers. That is, the defender attempts to recover the inputs given hidden state $h \in \mathcal{H}$. The defender network and the bipartite network are trained concurrently with the defender providing feedback regarding the efficacy of the privacy partitioning during each round and the bipartite network updating it's function $f_{\theta_d}:\mathcal{H}\rightarrow\mathcal{X}$ based on this information (see~\figref{partition:fig:model} (a)).

Suppose we would like to learn a deep learning model $f_\theta:\mathcal{X} \rightarrow  \mathcal{Y}$ with the training data $\mathcal{D} = \{ (x_i, y_i) \}_{i=1}^m$. According to our bipartite design, the deep learning model can be formulated as $f_\theta=f_{\theta_l} \circ f_{\theta_r} = f_{\theta_r}(f_{\theta_l}(\cdot))$, where $f_{\theta_l}: \mathcal{X}\rightarrow\mathcal{H}$ is the function mapping from input domain $\mathcal{X}$ to domain of the intermediate activation state $\mathcal{H}$ in local partition $\Theta_l$ and  $f_{\theta_r}: \mathcal{H}\rightarrow\mathcal{Y}$ is the function mapping from $\mathcal{H}$ to output domain $\mathcal{Y}$ in remote partition $\Theta_r$.

According to our design, the defender learns a mapping function $f_{\theta_d}: \mathcal{H} \rightarrow \mathcal{X}$ and its objective can be formulated as:

\begin{equation}
    \min_{\theta_{d}} \frac{1}{m}\sum_{i=1}^{m}d\big(x_i, f_{\theta_d}(f_{\theta_l}(x_i)) \big)
\end{equation}

where $d\big( \cdot, \cdot \big)$ is the privacy distance metric between the original input and the recovered input by the defender that measures how the original input and the recovered input differ in sensitive information of the original input. 

While training the defender, the model would leverage the defender's recovery performance as side information to better optimize its parameter: to make it harder for the attacker in the inference phase to recover input data as well as achieving original ``goal'' of the model:

\begin{equation} \label{partition:eq:defender-feedback}
    \min_{\theta} \frac{1}{m} \sum_{i=1}^{m} l\big(y_i, f_\theta(x_i)\big) - \lambda \cdot d\big(x_i, f_{\theta_d}(f_{\theta_l}(x_i))\big)
\end{equation}

where $ \theta= \{\theta_l, \theta_r\}$, $l\big(\cdot, \cdot \big)$ denotes the loss function for the original task and $\lambda$ is the defender weight. 

\subsection{Model Inference Phase}

When the model is properly deployed according to our bipartite topology design in model learning phase, the adversary wants to learn the best mapping function $f_{\theta_a}: \mathcal{H} \rightarrow \mathcal{X}$ among all the attacker architectures he devises to recover the input of the model using the dataset $\mathcal{D}^\prime = \{\hat{x_i}\}_{i=1}^n$, so that whenever new data comes out from the local computing node, it can recover the input. Its objective function can be formulated as:

\begin{equation}
    \min_{\theta_a \in \{\theta_{a_1}, \dotsc, \theta_{a_k}\}}\min_{\theta_{a}} \frac{1}{n}\sum_{i=1}^{n} d\big(\hat{x_i}, f_{\theta_a}(f_{\theta_l}(\hat{x_i}))\big)
\end{equation}

 Note the data set $\mathcal{D}^\prime$ that the attacker uses might be different from the model's training data. Attacker's data can be the data the attacker collected by himself, part of the training data set or the whole training data in the worst case. \figref{partition:fig:model} shows our framework in different phases.

\subsection{Benefits and Trade-offs}

The bipartite topology maps well the commonly occurring dual domain access control requirement (\eg server/client, public/private, host/guest) and extends well to domain-based access control in the IoT, fog computing, and edge computing paradigms. In the next section we evaluate the framework in terms model accuracy and protection strength to demonstrate these protections result in negligible degradation of performance (\ref{data:prop:performant}). The privacy partitioning framework is constructed using standard protocol-level mechanisms, thus, these privacy protections do not require processes that are outside of the deep learning tool-set (\ref{data:prop:learning-based}). 

It can be applied directly to centralized deep learning (\ref{data:prop:compatible}). It is also complementary to decentralized architectures and federated resource sharing model (\ref{data:prop:complementary}). Recent WPA2 Krack Attack~\cite{vanhoef2017key} demonstrates the importance of having a layered approach to security. DNNs are so prevalent that protection via protocol construction is useful. 

This approach does introduce more components during the model learning phase. The additional components (primarily the defender) result in an increase in required computing resources and time to learn a high-performance deep learning model. However, the increased model complexity is not a substantial usability or cost barrier when compared to the baseline cost of deploying a similar network with no protections. 
In the model inference phase, since we only deploy a portion of model layers in the local domain, it would require less computational resources in local domain and does not cause any overhead. Therefore, it is feasible for our framework to be deployed in the mobile devices or IoT devices.

\subsection{Hardening Privacy Partitioning with Defender Suites}

This depiction of the model learning phase (\figref{partition:fig:model} (a)) shows a single defender deployed at a single deep network partition. In practice, the privacy partitioning protections for a given deep network $f_\theta$ may be extended to include more than one defender loss function at a given privacy partitioning.
In other words, the defender $f_{\theta_d}$ can be extended to \emph{defender suite} $\mathcal{F}_{D}$=\{$f_{\theta_{d0}}$, $f_{\theta_{d1}}$, $\dotsc$, $f_{\theta_{dD}}$\}) at the cost of increasing computational complexity of the network, and increasing both time it takes to train. Including more defenders at a partition provides more robust privacy protections for the associated hidden state since the model can leverage "best" defender among all in the defender suite in the model learning phase, which can be formulated as

\begin{equation} 
\begin{split}
     \min_{\theta=(\theta_l, \theta_r)} & \frac{1}{m} \Big( \sum_{i=1}^{m} l\big(y_i, f_\theta(x_i)\big) \\
    &- \lambda \min_{\theta_d \in \{\theta_1, \dotsc, \theta_D\}} \min_{\theta_d} \sum_{i=1}^{m} d \big(x_i, f_{\theta_d}(f_{\theta_l}(x_i)) \big) \Big)
\end{split}
\end{equation}

However, solving this optimization problem is difficult due the complex architectures of the model and the defenders. In practice, the model provider may train multiple defenders individually and select choose the best defender with the recovered inputs most similar to the inputs, and update the parameters based on this defender in the model using iterative optimization method like Stochastic Gradient Descent (SGD). The whole model learning process with multiple defenders is shown in \algref{partition:alg:multi-defender}.

\begin{algorithm}
    \SetKwInOut{Input}{Input}
    \SetKwInOut{Output}{Output}
    \SetKwInOut{Initialize}{Initialize}
    
    \Input{Training set $\mathcal{D} = \{(x_i, y_i)\}_{i=1}^m$}
    \Output{Trained model $f_\theta$}
    
    \Initialize{model $\theta$ and defender suite $\{\theta_{d1}, \dotsc, \theta_{dD}\}\;$}
    
    \For{$t \in [T]$}{
         \For{\textup{each mini-batch $\{x_i, y_i\}_{i=1}^B \in \mathcal{D}$ }}{
        1. \textup{Update the defender suite parameters $ \theta_d \in \{\theta_{d1}, \dotsc, \theta_{dD}\}$ via SGD using the gradient}
        \begin{equation}
        \nonumber
            \nabla_{\theta_{d}} \frac{1}{m}\sum_{i=1}^{m}d\big(x_i, f_{\theta_d}(f_{\theta_l}(x_i)) \big)
        \end{equation}
        
        2. \textup{Choose the best defender via}
        \begin{equation}
        \nonumber
        \theta_d = \argmin_{\theta_d \in \{\theta_{d1}, \dotsc, \theta_{dD}\}}  \frac{1}{B}\sum_{i=1}^{B}d\big(x_i, f_{\theta_d}(f_{\theta_l}(x_i)) \big)
         \end{equation}
        3. \textup{Update the model parameters $\theta$  via SGD using the graident}
        \begin{equation}
        \nonumber
        \nabla_{\theta} \frac{1}{B} \sum_{i=1}^{B} l\big(y_i, f_\theta(x_i)\big) - \lambda \cdot d\big(x_i, f_{\theta_d}(f_{\theta_l}(x_i))\big)
        \end{equation}
        } 
    }
    \Return $\theta$\;
    
    \caption{Privacy Partition with Multiple Defenders}
    \label{partition:alg:multi-defender}
\end{algorithm}

\subsection{Performance and Privacy Controls with Partition Suites} 

Another strategy for hardening the privacy partitioning framework is to extend a bipartite network \{$\Theta_{l}$, $\Theta_{r}$\}) with one privacy partitioning $\{\mathcal{L}_{i},\mathcal{L}_{i+1}\}$ into a partition suite ($\{\mathcal{L}_{i},\mathcal{L}_{i+1}\}, \{\mathcal{L}_{j},\mathcal{L}_{j+1}\},  \hdots, \{\mathcal{L}_{m},\mathcal{L}_{m+1}\}$). Integration can occur either by adding multiple partitions to a single deep network or by training a new deep network for each partition in the suite. Simultaneously learning more than one privacy partitioning provides a more complete protection surface and more deployment configuration options at the cost of additional training overhead. The additional configurations enable flexible controls that can be adjusted in near real-time to suit changing requirements of contextual data streams and dynamic end-user privacy requirements.

\subsection{Continuous Learning with Locked Local Partitions} 

Although the required steps for adding a privacy partitioning occur during the model learning phase, the result is a bipartite deep network with built-in protections for the inference phase. For this reason, this section focuses primarily on putting the protections in place during the learning phase whereas as the threat model (\secref{partition:sec:threat-model}) and evaluation (\secref{partition:sec:experiments}) focus on the inference phase.

We now describe a process by which a privacy partitioned bipartite deep network may be updated online to incorporate new data. We will denote this phase as the \emph{online learning phase} to distinguish it from the learning phase described thus far (\ie the initial learning phase). The process for supporting online updating of the remote partition occurs as follows:

\begin{enumerate}
    
    \item initial learning phase:
    \begin{enumerate}
        \item select reference network $\Theta$ 
        \item select privacy partitioning point $\{\mathcal{L}_{i},\mathcal{L}_{i+1}\}$
        \item learn bipartite network $\{\Theta_{l},\Theta_{r}\}$
    \end{enumerate}

    \item online learning phase:
    \begin{enumerate}
        \item lock local partition $\Theta_{l}$
        \item securely generate updates $f_{\theta_l}:\mathcal{X}\rightarrow\mathcal{H}$
        \item update remote partition $f_{\theta_r}:\mathcal{H}\rightarrow\mathcal{Y}$
        \item validate and test update $f_{\theta^*}:\mathcal{X}\rightarrow\mathcal{Y}$
        \item deploy bipartite network update $\{\Theta_{l},\Theta_{r^*}\}$
    \end{enumerate}

\end{enumerate}

The introduction of an online learning phase, occurring after the initial model is learned, has the primary benefit of completely removing any requirement to conduct data collection and curation tasks in the remote computing context. Our construction thus far assumes an initial learning stage where a single administrative domain requires access the entire topology as well as training data. During the online learning and the subsequent inference phases, the protections of privacy partitioning extend to training, validating, and testing processes. Thus, there is no need for a specialized single admin learning phase after the initial learning phase used to generate the local partition. Further, a remote admin and local admin can negotiate an online update without having to grant training data access to the remote computing context or requiring local computing context to manage the entire process.

\section{Experiments}\label{partition:sec:experiments}
%Experiments

In this section, we experimentally demonstrate that the proposed framework can successfully defend against the attackers described in the threat model. In the experiment setting, we assume that all attackers have access to the full training dataset. This assumption makes it more challenging for our proposed solution to protect the data privacy since the attacker can have the exact data distribution trained by the model. Also, in practice, this can sometimes happen when the model provider is malicious or some hackers can gain unauthorized access to the training data. 

We first discuss the metrics we used in our experiments to evaluate the effectiveness of our proposed methodology in the domain of for image privacy (\secref{partition:sec:exp-eval-metrics}). Then we evaluate the privacy partitioning adaptations to a three-layer neural network constructed using the MNIST dataset serving as a toy problem (\secref{partition:sec:exp-digit-id}). The second set of experiments evaluates the practicality of the privacy partitioning in protecting the sensitive information of input data using the LFW dataset (\secref{partition:sec:exp-face-id}). At last, we validate the effectiveness of our framework on the state-of-the-art very deep CNN using the CIFAR-10 dataset (\secref{partition:sec:exp-cifar10}). 

\subsection{Evaluation Metrics} \label{partition:sec:exp-eval-metrics}
How to judge the privacy leakage in images has long remained challenging in the field of computer vision. Different users might have different privacy requirements in different applications. For example, some users might not want to disclose the gender/age information of portrait images while he/she would still want to upload portrait images for deep learning services. 
In general cases, computer vision researchers use indistinguishable metrics to evaluate the quality of images compared to the benchmark images. Many algorithms and metrics~\cite{wang2004image, wang2003multiscale, zhang2011fsim, mantiuk2011hdr} have been proposed to quantify the human perceptual capability over images. 
Many Studies based on user studies and statistical evaluations have been proved these metrics like Structural Similarity (SSIM) index are highly consistent with human perceptual capability on different image quality measurement datasets~\cite{sheikh2006statistical, ponomarenko2009tid2008, larson2010most, ponomarenko2015image}. 
In the experiments below, we will use a combination of metrics discussed below to measure the indistinguishability of the recovered images by the attacker and as the proxy of the image privacy metrics.

\textbf{MSE}: The mean squared error (MSE) measures the per-pixel $l$-2 Euclidean distance between two images.  It is usually used to measure the quality of image reconstruction. In the context of image or video compression, it is often used as an approximation to human perception of reconstruction quality. However, the MSE is insufficient to assess highly-structured images since it assumes pixel-wise independence. For example, blurring an image can result in small $l$-2 changes but large perceptual changes. If MSE $=0$, two images are identical. 

\textbf{SSIM}: The structural similarity index (SSIM) improves on MSE  by assuming pixel-wise dependence~\cite{wang2004image}.  It is computed on a sliding window to capture the structural information between two images. It ranges from $[-1, 1]$, where SSIM $=1$ indicates that the two images are identical. 
SSIM is highly consistent with human perceptual capability so that image obfuscation techniques such as pixelation and blurring would result in smaller SSIM values.

\textbf{DPD}: Recently, studies have been shown that internal activations of deep convolutional networks trained on image inferences are surprisingly useful to capture “perceptual loss” that correspond to human visual perception~\cite{dosovitskiy2016generating, zhang2018unreasonable} especially in terms of perceptual spatial ambiguities.  In our experiment, we use the deep perceptual distance (DPD) proposed in~\cite{zhang2018unreasonable} to evaluate the perceptual indistinguishability of recovered images as a complementary metric of SSIM and MSE. 

\textbf{Reprint accuracy}: The reprint accuracy is a measure of the model classification accuracy using recovered inputs by the attacker (\ie the classification accuracy of $f:\mathcal{H}\rightarrow\mathcal{X}\rightarrow\mathcal{Y}$).  It can be a proxy metric to measure how much information loss of the recovered inputs compared to the original inputs. 

\begin{figure*}[ht!]
\begin{mdframed}[
    tikzsetting={align=center,},
    innerrightmargin=25pt, innerleftmargin=25pt, innerbottommargin=5pt, innertopmargin=5pt, topline=false,
  rightline=false,
  leftline=false,
  bottomline=false
]
    
    \centering
    
    \psfrag{a}[c][c][0.9]{\makecell{original images\\(a)}}
    \psfrag{b}[c][c][0.9]{\makecell{no defender\\(b)}}
    \psfrag{c}[c][c][0.9]{\makecell{has defender\\(c)}}
    
    \includegraphics[width=6.2in]{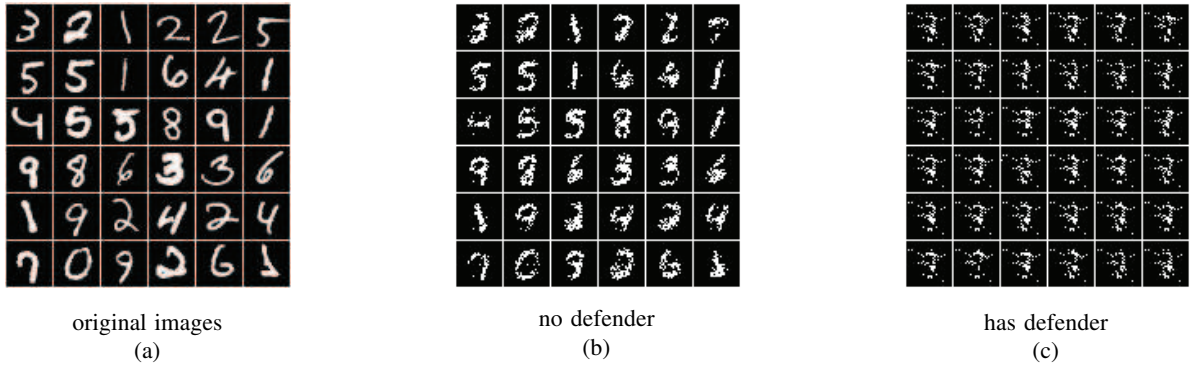}
    \caption{{\bf Handwritten Digit Image Recovery:} This figure compares the input images (a) to a \texttt{handwriting-to-digit} classifier network to the images recovered by the two-layer ReLU-based neural network from this classifier network's intermediate layer output when the privacy partition \emph{is not} applied (b) and when the privacy partition \emph{is} applied (c). The input to the recovery network consists of the intermediate layer output passed between the local domain and the remote domain of the classifier network. The MSE for the images in (b) and (c) are $\approx$0.07 and $\approx$0.11 (SSIM $\approx$0.55 and $\approx$0.35), respectively.}
    \label{partition:fig:mnist-recover}
    
\end{mdframed}
\end{figure*}

\subsection{Preliminary Validation of Privacy Partitioning using MNIST Dataset} \label{partition:sec:exp-digit-id}

As a preliminary step, we test the effectiveness of our method with a simple benchmark dataset. We choose one of the most commonly used Image benchmark datasets MNIST to start our tests. Experiments on MNIST show that our solution can make it much harder for attackers to recover input data during model inference phase. Since MNIST contains less sensitive data, we then run experiments on face image to show that our approach can protect private information such as gender while still provide high classification accuracy of face identification, see details in Section~\ref{partition:sec:exp-face-id}.  

We use the MNIST dataset~\cite{lecun1998mnist} to learn the \texttt{handwriting-to-digit} classifier network. MNIST is a benchmark computer vision dataset containing 60,000 gray-scale images of handwritten digits (50,000 training and 10,000 testing). We use a three-layer ReLU-based fully-connected feed-forward neural network with 800 hidden units in each layer as our model. We refer to this model as \texttt{handwriting-to-digit} classifier network. In the model inference phase, the first two layers are deployed on the local side and the last layer is deployed in the remote side. The attacker then leverages the output of the local layers to recover input images.

In all of our experiment settings for MNIST, the model and the defender(s) are trained using the Adam optimization algorithm \cite{kingma2014adam}. We set the learning rate of the model and the defender at 0.0001 and 0.001, respectively. We train the model and the defender for 500 epochs with a batch size of 32. we also use the dropout technique \cite{srivastava2014dropout} with a drop probability of 0.1 to prevent over-fitting.

We use the SSIM as the distance metric (defined in \eqnref{partition:eq:defender-feedback}).
We also use the Mean Square Error (MSE) loss as the loss function for training the attacker models. We choose MSE because it is typically used as the loss function in computer vision task. Each attacker model is saved when the MSE loss achieves its minimum during training.

We evaluate the quality of attack using both MSE and SSIM between the recovered images by all attackers and the input images in the test set.

 \textit{(1) How do different defender weights affect the performance of attackers?}\hfill

First, we consider the case where the model provider uses only one defender in the model learning phase. We set the defender model as a two-layer ReLU-based neural network. The attacker uses different network architectures to recover the input images in the model inference phase. In order to try out the different possible model architectures, we choose the attacker models based on different combination of hyperparameters, such as activation function, number of hidden layers, number of neuron per layer, etc. We choose 8 attacker models in total. The details of chosen attacker models can be found in~\appref{partition:appendix:mnist}.

We set the defender weight $\lambda$ (see in \eqnref{partition:eq:defender-feedback}) to be $0, 100, 200, 300, 400, 500$ to see the corresponding model accuracy and the quality of attacker recovered images (SSIM and MSE). The results are shown in~\tabref{partition:tab:mnist-n-v-1}.

\begin{table*}[ht]
\begin{footnotesize}
\centering
  \ra{1.5} % manually adjust the line height of table rows

    \caption{{\bf MNIST Model with Multiple Attackers Versus a Single Defender}}
    \label{partition:tab:mnist-n-v-1}
    
    \begin{tabular}{|l|l|r|r|r|r|r|r|}
        \hline
        
        \multicolumn{2}{|l|}{{\bf Defender Weight}} & 0 & 100 & 200 & 300 & 400  & 500  \\\hline
        \multicolumn{2}{|l|}{{\bf Model Accuracy}} & 98.4\%	& 98.2\%	& 98.2\%	& 98.1\%	& 98.1\%	& 98.1\%  \\\hline\hline
       
       \multirow{8}{*}{{\bf MSE/SSIM}} 
        & Attacker $\circled{1}$    & 0.070/0.546 &	0.072/0.534 &	0.079/0.500 &	0.081/0.495 &	0.086/0.466 &	0.116/0.347 \\
        \cline{2-8}
        & Attacker $\circled{2}$    & 0.072/0.524 &	0.068/0.553 & 0.080/0.500 &	0.082/0.491 &	0.093/0.449 &	0.117/0.361  \\
        \cline{2-8}
        & Attacker $\circled{3}$    & 0.016/0.834 &	0.018/0.823 &	0.021/0.807 &	0.021/0.797 &	0.021/0.818 &	0.023/0.792  \\
        \cline{2-8}
        & Attacker $\circled{4}$    & 0.022/0.783 &	0.021/0.794 &	0.025/0.755 &	0.025/0.765 &	0.023/0.787 &	0.025/0.760  \\
        \cline{2-8}
        & Attacker $\circled{5}$    & 0.066/0.552 & 0.062/0.587 &	0.071/0.533 & 0.076/0.517 & 0.076/0.517 &	0.087/0.457  \\
        \cline{2-8}
        & Attacker $\circled{6}$    & 0.074/0.514 &	0.070/0.543 &	0.087/0.457 &	0.097/0.420 &	0.092/0.448 &	0.131/0.298  \\
        \cline{2-8}
        & Attacker $\circled{7}$    & 0.070/0.527 &	0.063/0.575 &	0.071/0.531 &	0.070/0.532 &	0.081/0.470 &	0.091/0.443  \\
        \cline{2-8}
        & Attacker $\circled{8}$    & 0.032/0.734 &	0.037/0.713 &	0.046/0.659 &	0.049/0.637 &	0.048/0.640 &	0.061/0.558  \\\hline
        % \hline
        
        % \multirow{8}{*}{{\bf SSIM}} 
        % & Attacker $\circled{1}$    & 0.546 &	0.534 &	0.500 &	0.495 &	0.466 &	0.347   \\
        % \cline{2-8}
        % & Attacker $\circled{2}$    & 0.524 &	0.553 &	0.500 &	0.491 &	0.449 &	0.361  \\
        % \cline{2-8}
        % & Attacker $\circled{3}$    & 0.834 &	0.823 &	0.807 &	0.797 &	0.818 &	0.792  \\
        % \cline{2-8}
        % & Attacker $\circled{4}$    & 0.783 &	0.794 &	0.755 &	0.765 &	0.787 &	0.760\\
        % \cline{2-8}
        % & Attacker $\circled{5}$    & 0.552 &	0.587 &	0.533 &	0.517 &	0.517 & 0.457  \\
        % \cline{2-8}
        % & Attacker $\circled{6}$    & 0.514 &	0.543 &	0.457 &	0.420 &	0.448 &	0.298  \\
        % \cline{2-8}
        % & Attacker $\circled{7}$    & 0.527 &	0.575 &	0.531 &	0.532 &	0.470 &	0.443  \\
        % \cline{2-8}
        % & Attacker $\circled{8}$    & 0.734 &	0.713 &	0.659 &	0.637 &	0.640 &	0.558  \\
        % \hline

    \end{tabular}
     \\ [2ex] %\justify
     \begin{center}
    The overall trend is a reduction in attacker recovery accuracy as defender weights are increased.
    \end{center}
\end{footnotesize}
\end{table*}

\textbf{Result Analysis:} In \tabref{partition:tab:mnist-n-v-1}, we can clearly see that for each attacker model, the overall trend of MSE increases (SSIM decreases) with the increase of the defender weight, and in all cases the inference accuracy still remains at the high level ($>98\%$). The results demonstrate the effectiveness of our framework: adding defender in our framework make it harder for the attacker to recover input images while maintaining the model inference accuracy. \figref{partition:fig:mnist-recover} shows MNIST images recovered by the two-layer ReLU-based neural network with the defender present and without the defender present. We can clearly see that with the defender present, the recovered images by the attacker network is harder to recognize.

For all attacker models, the defender works to some extent compared with the case of no defender. However, for some attacker models such as Attacker $\circled{3}$ and Attacker $\circled{4}$, even though a defender is present when training the model, the MSEs is still low (SSIM is still high). It indicates that the single chosen defender architecture does not harden these two attacker models enough.

 \textit{(2) How do multiple defenders affect the performance of multiple attackers?}\hfill

We have shown that the chosen defender architecture in the previous experiment cannot perfectly defend against multiple different types of attackers. Next, we consider extending the experiment by adding multiple defenders during the model learning phase to defend against multiple attackers again. We carefully select the 4 defenders based on our previous experiments result: we choose those defender models which perform the best (in both SSIM and MSE) as attacker models in the previous experiment. The details of the chosen defenders are in \appref{partition:appendix:mnist}.

As for the attacker networks, we use the same 8 attackers used in the previous experiment for comparison to see how the case of multiple defenders improves from that of a single defender. We choose the defender weight $\lambda$ as 200 for comparison. Experiment results are shown in \tabref{partition:tab:mnist-n-v-n}.

\begin{table}[ht]
\centering
  \ra{1.4} % manually adjust the line height of table rows

\caption{{\bf MNIST Model with Multiple Attacker versus Multiple Defenders} %Training the model with multiple defenders hardens against each attacker model, reducing their recovery potential.
}
\label{partition:tab:mnist-n-v-n}

\begin{tabular}{|l|l|r|r|r|}
\hline
\multicolumn{2}{|l|}{{\bf Defenders Present}}                                              & {\bf \makecell{No\\Defender}}&  {\bf \makecell{Single\\Defender}} & {\bf \makecell{Multiple\\Defenders}} \\ \hline
\multicolumn{2}{|l|}{{\bf Model Accuracy}}                                           & 98.4\%  & 98.2\%    & 98.0\%        \\ \hline\hline
\multicolumn{1}{|c|}{\multirow{8}{*}{{\bf MSE}}} & Attacker $\circled{1}$          & 0.070 & 0.079 & 0.215         \\ \cline{2-5} 
\multicolumn{1}{|c|}{}                     & Attacker $\circled{2}$ & 0.072 & 0.080     & 0.209         \\ \cline{2-5} 
\multicolumn{1}{|c|}{}                     & Attacker $\circled{3}$ & 0.016 & 0.021     & 0.070         \\ \cline{2-5} 
\multicolumn{1}{|c|}{}                     & Attacker $\circled{4}$ & 0.022  & 0.025    & 0.073         \\ \cline{2-5} 
\multicolumn{1}{|c|}{}                     & Attacker $\circled{5}$ & 0.066 & 0.071     & 0.195         \\ \cline{2-5} 
\multicolumn{1}{|c|}{}                     & Attacker $\circled{6}$ & 0.074 & 0.087     & 0.202         \\ \cline{2-5} 
\multicolumn{1}{|c|}{}                     & Attacker $\circled{7}$ & 0.070 & 0.071     & 0.168         \\ \cline{2-5} 
\multicolumn{1}{|c|}{}                     & Attacker $\circled{8}$ & 0.032 & 0.046     & 0.192         \\ \hline\hline
\multirow{8}{*}{{\bf SSIM}}                      & Attacker $\circled{1}$ & 0.530 & 0.500      & 0.076         \\ \cline{2-5} 
& Attacker $\circled{2}$ & 0.540  & 0.500    & 0.098         \\ \cline{2-5} 
& Attacker $\circled{3}$    & 0.841 & 0.807     & 0.500         \\ \cline{2-5} 
& Attacker $\circled{4}$ & 0.820  &   0.755  & 0.494         \\ \cline{2-5} 
& Attacker $\circled{5}$ & 0.543  &  0.533   & 0.194         \\ \cline{2-5} 
& Attacker $\circled{6}$ & 0.518  &  0.457   & 0.117         \\ \cline{2-5} 
& Attacker $\circled{7}$ & 0.568  &  0.531   & 0.205         \\ \cline{2-5} 
& Attacker $\circled{8}$ & 0.722  &  0.659   & 0.160         \\ \hline
\end{tabular}

\end{table}

\textbf{Result Analysis:} In \tabref{partition:tab:mnist-n-v-n}, we can clearly see that the performance of each attacker model degrades by a large extent when there are multiple defenders present, compared to the performance of each attacker model when there is no or only one defender present. This is because by training multiple defenders in the model learning phase, the model can choose the best defender in each step that has the best recover-ability to optimize its parameters.
This will harden the chance that the attacker models recover the input images, and decrease indistinguishability of the recovered images by the attacker.

However, adding more defenders during training always mean more computing resources in practice. Therefore, we should strike a balance between the availability of computing resources or try to find a better and more representative defender model architecture which greatly decrease the inevitability of local layers of the model given the data distribution.

\subsection{Protecting Private Information in Face Dataset}\label{partition:sec:exp-face-id}

\begin{figure*}[!ht]
\begin{mdframed}[
    tikzsetting={align=center, draw=black, thick, align=center},
    innertopmargin=10pt,innerrightmargin=10pt,innerleftmargin=10pt,innerbottommargin=10pt, topline=false,
  rightline=false,
  leftline=false,
  bottomline=false
]
    \centering
    
    % defender configurations
    \psfrag{a}[c][c][0.9]{\makecell{has\\defender}}
    \psfrag{b}[c][c][0.9]{\makecell{no\\defender}}
    
    % layer configurations
    \psfrag{c}[c][c][0.9]{$\mathcal{L}_3$ (\texttt{pool1})}
    \psfrag{d}[c][c][0.9]{$\mathcal{L}_6$ (\texttt{pool2})}
    \psfrag{e}[c][c][0.9]{$\mathcal{L}_9$ (\texttt{pool3})}
    
    % axis labels
    \psfrag{f}[c][c][0.9]{\rotatebox{90}{defenders}}
    \psfrag{g}[c][c][0.9]{local layers}
    
    % original images
    \psfrag{h}[c][c][0.9]{original images}

    \includegraphics[width=6.2in]{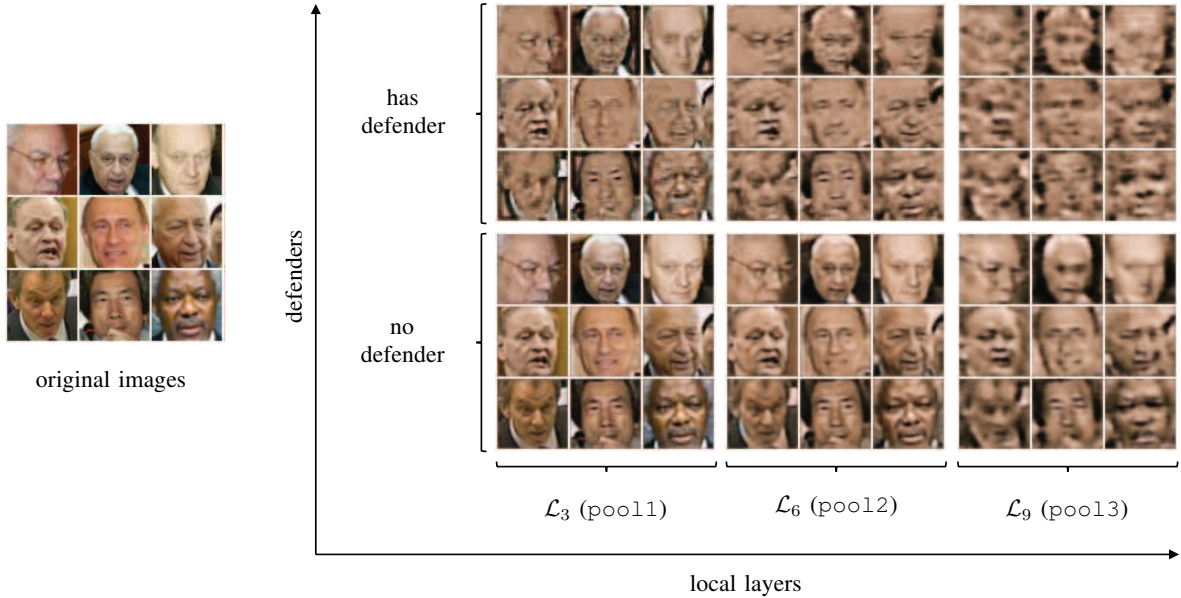}
    \caption{{\bf Faces Photo Recovery:} This figure compares nine sample images recovered by six configurations (2 defender configurations by 3 partition configurations) of classifier network \texttt{photo-to-id}. Note that the recovery error both increase as a defender is applied to the privacy partition (bottom-to-top trend) and as more layers are included in the local partition $\Theta_l$ (left-to-right trend) (refer to \figref{partition:fig:face-cifar-results} for the full results of the \texttt{photo-to-id} experiments).}
    \label{partition:fig:face-id-recover}
    
\end{mdframed}    
\end{figure*}

We use the Labeled Faces in the Wild (LFW) dataset \cite{huang2008labeled} to learn the face recognition classifier network \texttt{photo-to-id}. The LFW dataset contains 13,233 images of faces collected from the web and each face was labeled with the name of the person in the picture. There are 1,680 people who have at least two distinct images. The size of the images is $250 \times 250$. We only retain images of people that have at least 30 different pictures and re-scale the images to $64 \times 64$. This is a typical input pipeline of face recognition. After filtering, there are 2,370 images left with 34 subjects in total. We split $80\%$ as the training set and the rest $20\%$ as the testing set in a stratified fashion.

We use the CNN as the classifier. The details of the CNN classifier architecture is in \appref{partition:appendix:id}.

During the model learning phase, the defender architecture we add for training the classifier is similar to the ``reversed'' version of the local layers.
The defender ``reverses'' the model operations to get the input images from the output of local layers (\eg to reverse the input of convolution operation or a pooling operation, the defender perform a deconvolution operation on the hidden-layer features~\cite{zeiler2010deconvolutional}). This type of architecture resembles the design strategy of a convolutional autoencoder~\cite{hinton2006reducing, erhan2010does}.

 The model and the defender are trained using SGD algorithm with momentum $0.9$ and initial learning rate $0.01$. The learning rate decays by a factor of $0.1$ every $100$ epochs, and there are $250$ epochs in total. $l$-$2$ regularization is also applied to prevent over-fitting. The defender's weight $\lambda$ is set to be $0.1$ so that the model inference performance is at an acceptable level.

In the model inference phase, we choose the attacker architecture based on decoder designs the autoencoder~\cite{hinton2006reducing, erhan2010does}. We choose three different types of attacker models to recover the input images: (1) the attacker model with deconvolution layer as its main component to recover the input images;
(2) the attacker model that use fully-connected layers as its main component;
(3) the attacker model that use sparse fully-connected layers as its main component (we introduce the sparsity of the network by dropout techniques~\cite{srivastava2014dropout}).
We choose these types of attackers since they are the most commonly used decoder architecture in auto-encoder designs~\cite{hinton2006reducing} and they cover the most commonly used operations in deep neural networks in the area of computer vision.
The detailed architectures of the attacker models are in \appref{partition:appendix:id}.

 \textit{(1) Where should we place the privacy partitioning on our model}\hfill

To strike a balance between model utility and data privacy, we should choose the intermediate layer whose hidden state is difficult for the attacker to recover the input images while also maintaining the model classification accuracy. A question arises that which layer we should choose for the privacy partitioning. 

To answer the question, we choose different layers for the privacy partitionings in the model learning phase.
% \figref{partition:fig:face-id-topology} shows the different partitions for our experiment. 
We choose the output of the first, second and third pooling layer (denoted by \texttt{pool1}, \texttt{pool2}, \texttt{pool3}) for privacy partitionings (see \appref{partition:appendix:id} for details). 
We choose these layers because the outputs of these layers represent different levels of abstraction of features in the model in the feature extraction module in the CNNs~\cite{zeiler2014visualizing}.

\begin{figure*}[!ht]
\begin{mdframed}[
    tikzsetting={align=center,},
    innerrightmargin=5pt, innerleftmargin=5pt, innerbottommargin=5pt, innertopmargin=5pt, topline=false,
  rightline=false,
  leftline=false,
  bottomline=false
]

\centering
\begin{subfigure}{0.15\textwidth}
\begin{tikzpicture}[scale=0.38]
    \centering\Large
    \begin{axis}[xbar,enlargelimits=0.20, bar width=1.5,
    ytick={5,10,15},
    yticklabels={$H_0$,$H_1$,$H_2$},
    ylabel=Partition Point,
    xlabel=Model Accuracy (\%),
    xmajorgrids,
    xmin=70, xmax=100, xtick={70,80,90,100},
    legend style={at={(0.5,1.12)},anchor=north},
    legend columns=-1,],

    \addplot[pattern=north west lines, pattern color=gray] % has defender
    [draw=black] 
    coordinates {(88.19,5) (92.19,10) (90.72,15)};
    % 88.19, 92.19, 90.72
    
    \addplot[fill=black] % no defender
    [draw=black] 
    coordinates {(92.55,5) (92.55,10) (92.55, 15)};
    % 92.55

    \legend{has defender, no defender}
    
    \end{axis}
\end{tikzpicture}
\caption{} %a
\centering
\end{subfigure}
\hfil
\begin{subfigure}{0.15\textwidth}
\begin{tikzpicture}[scale=0.38]
    \centering\Large
    \begin{axis}[xbar,enlargelimits=0.20,bar width=1.5,enlarge x limits=0,
    ytick={5,10,15},
    ylabel=Partition Point,
    xlabel=Reprint Accuracy (\%),
    yticklabels={$H_0$,$H_1$,$H_2$,$H_{3}$,$H_{4}$},
    xmajorgrids,xmin=0, xmax=100, xtick={0,20,40,60,80,100},
    legend style={at={(0.5,1.12)},anchor=north},
    legend columns=-1,],
    \addplot[pattern=north west lines, pattern color=gray] % has defender
    [draw=black] 
    coordinates {(71.308, 5) (80.802,10) (59.705,15)};
    % 71.308, 80.802, 59.705
    \addplot[fill=black] % no defender
    [draw=black] 
    coordinates {(91.350, 5) (89.241,10) (81.013,15)};
    %91.350, 89.241, 81.013
    \legend{has defender, no defender}
    \end{axis}
\end{tikzpicture}
\caption{} %b
\end{subfigure}
\hfil
\begin{subfigure}{0.15\textwidth}
\begin{tikzpicture}[scale=0.38]
    \centering\Large
    \begin{axis}[xbar,enlargelimits=0.20,bar width=1.5,
    ytick={5,10,15},
    ylabel=Partition Point,
    xlabel=SSIM,
    xtick={0.2, 0.4,0.6,0.8,1.0},
    yticklabels={$H_0$,$H_1$,$H_2$},xmajorgrids,
    legend style={at={(0.5,1.12)},anchor=north},
    legend columns=-1],
    
    \addplot[pattern=north west lines, pattern color=gray] % has defender
    [draw=black] 
    coordinates {(0.6144, 5) (0.5184,10) (0.4241,15)};
    % 0.6144, 0.5184, 0.4241
    \addplot[fill=black] % no defender
    [draw=black] 
    coordinates {(0.9198, 5) (0.8275,10) (0.6035,15)};
    % 0.9198, 0.8275, 0.6035
    \legend{has defender, no defender}
    
    \end{axis}
\end{tikzpicture}
\caption{} %c
\end{subfigure}
\hfil
\begin{subfigure}{0.15\textwidth}
\begin{tikzpicture}[scale=0.38]
    \centering\Large
    \begin{axis}[xbar,enlargelimits=0.20, bar width=1.5,
    ytick={5,10,15},
    ylabel=Partition Point,
    xlabel=MSE,xmin=0, xmax=2.5,enlarge x limits=0,
    yticklabels={$H_0$,$H_1$,$H_2$},
    xmajorgrids,
    legend style={at={(0.5,1.12)},anchor=north},
    legend columns=-1,],
    \addplot[pattern=north west lines, pattern color=gray] % has defender
    [draw=black] 
    coordinates {(1.0295, 5) (1.9334,10) (2.2982,15)};
    %787.543572, 1479.015676, 1758.157714
    % need to be normalized / (255*3)
    \addplot[fill=black] % no defender
    [draw=black] 
    coordinates {(0.2175, 5) (0.4819,10) (1.0647,15)};
    % 166.395242, 368.669703, 814.472241
    % need to be normalized / (255*3)

    \legend{has defender, no defender}
    
    \end{axis}
\end{tikzpicture}
\caption{} %d
\end{subfigure}
\hfil
\begin{subfigure}{0.15\textwidth}
\begin{tikzpicture}[scale=0.38]
    \centering\Large
    \begin{axis}[xbar,enlargelimits=0.20,bar width=1.5,
    ytick={5,10,15},
    ylabel=Partition Point,
    xlabel=DPD,
    yticklabels={$H_0$,$H_1$,$H_2$,$H_{3}$,$H_{4}$},
    xmajorgrids,
    legend style={at={(0.5,1.12)},anchor=north},
    legend columns=-1,],

    \addplot[pattern=north west lines, pattern color=gray] % has defender
    [draw=black] 
    coordinates {(0.1518,5) (0.2485,10) (0.3898,15)};
    %0.1518, 0.2485, 0.3898
    \addplot[fill=black] % no defender
    [draw=black] 
    coordinates {(0.0673, 5) (0.1577,10) (0.2955,15)};
    % 0.0673, 0.1577, 0.2955

    \legend{has defender, no defender}
    
    \end{axis}
\end{tikzpicture}
\caption{} %e
\end{subfigure}

\begin{subfigure}{0.15\textwidth}
\begin{tikzpicture}[scale=0.38]
    \centering\Large
    \begin{axis}[xbar,enlargelimits=0.20,bar width=1.5,
    ytick={5,10,15},
    ylabel=Partition Point,
    xlabel=Model Accuracy (\%),
    yticklabels={$H_0$,$H_1$,$H_2$},
    xmajorgrids,
    xmin=70, xmax=100, xtick={70,80,90,100},
    legend style={at={(0.5,1.12)},anchor=north},
    legend columns=-1,],
    
    \addplot[pattern=north west lines, pattern color=gray] % has defender
    [draw=black] 
    coordinates {(91.48,5) (92.11,10) (92.88,15)};
    %91.48, 92.11, 92.88
    \addplot[fill=black] % no defender
    [draw=black] 
    coordinates {(93.10, 5) (93.10,10) (93.10,15)};
    % 93.10
    \legend{has defender, no defender}
    
    \end{axis}
\end{tikzpicture}
\caption{} %f
\end{subfigure}
\hfil
\begin{subfigure}{0.15\textwidth}
\begin{tikzpicture}[scale=0.38]
    \centering\Large
    \begin{axis}[xbar,enlargelimits=0.20,bar width=1.5,enlarge x limits=0,
    ytick={5,10,15},
    ylabel=Partition Point,
    xlabel=Reprint Accuracy (\%),
    yticklabels={$H_0$,$H_1$,$H_2$,$H_{3}$,$H_{4}$},
    xmajorgrids,xmin=0, xmax=100, xtick={0,20,40,60,80,100},
    legend style={at={(0.5,1.12)},anchor=north},
    legend columns=-1,],

    \addplot[pattern=north west lines, pattern color=gray] % has defender
    [draw=black] 
    coordinates {(84.01, 5) (36.77, 10) ( 20.95, 15)};
    %84.01, 36.77, 20.95
    \addplot[fill=black] % no defender
    [draw=black] 
    coordinates {(91.25, 5) (76.99, 10) (30.05, 15)};
    %91.25,76.99, 30.05

    \legend{has defender, no defender}
    
    \end{axis}
\end{tikzpicture}
\caption{} %g
\end{subfigure}
\hfil
\begin{subfigure}{0.15\textwidth}
\begin{tikzpicture}[scale=0.38]
    \centering\Large
    \begin{axis}[xbar,enlargelimits=0.20, bar width=1.5,
    ytick={5,10,15},
    ylabel=Partition Point,
    xlabel=SSIM,
    yticklabels={$H_0$,$H_1$,$H_2$,$H_{3}$,$H_{4}$},
    xmajorgrids,
    legend style={at={(0.5,1.12)},anchor=north},
    legend columns=-1,],

    \addplot[pattern=north west lines, pattern color=gray] % has defender
    [draw=black] 
    coordinates {(0.8242, 5) (0.357,10) (0.2699,15)};
% 0.8242, 0.357, 0.2699
    
    \addplot[fill=black] % no defender
    [draw=black] 
    coordinates {(0.9507,5) (0.7655,10) (0.344, 15)};
    % 0.9507, 0.7655, 0.344

    \legend{has defender, no defender}
    
    \end{axis}
\end{tikzpicture}
\caption{} %h
\end{subfigure}
\hfil
\begin{subfigure}{0.15\textwidth}
\begin{tikzpicture}[scale=0.38]
    \centering\Large
    \begin{axis}[xbar,enlargelimits=0.20,bar width=1.5,enlarge x limits=0,xmin=0,xmax=5,
    ytick={5,10,15},
    ylabel=Partition Point,
    xlabel=MSE,
    yticklabels={$H_0$,$H_1$,$H_2$},
    xmajorgrids,
    legend style={at={(0.5,1.12)},anchor=north},
    legend columns=-1,],
    \addplot[pattern=north west lines, pattern color=gray] % has defender
    [draw=black] 
    coordinates {(0.62445,5) (4.04748,10) (4.6136, 15)};
    %477.70755, 3096.325928, 3529.439941
    \addplot[fill=black] % no defender
    [draw=black] 
    coordinates {(0.30439, 5) (1.01135,10) (3.2709, 15)};
    %232.855148, 773.681702, 2502.236816
    \legend{has defender, no defender}
    
    \end{axis}
\end{tikzpicture}
\caption{} %i
\end{subfigure}
\hfil
\begin{subfigure}{0.15\textwidth}
\begin{tikzpicture}[scale=0.38]
    \centering\Large
    \begin{axis}[xbar,enlargelimits=0.20, bar width=1.5,enlarge x limits=0,xmin=0,xmax=0.25,
    ytick={5,10,15},
    ylabel=Partition Point,
    xlabel=DPD,
    yticklabels={$H_0$,$H_1$,$H_2$},
    xmajorgrids,
    legend style={at={(0.5,1.12)},anchor=north},
    legend columns=-1,],

    \addplot[pattern=north west lines, pattern color=gray] % has defender
    [draw=black] 
    coordinates {(0.0193,5) (0.1135,10) (0.2168, 15)};
    % 0.0193, 0.1135, 0.2168
    
    \addplot[fill=black] % no defender
    [draw=black] 
    coordinates {(0.0092, 5) (0.0468,10) (0.2029,15)};
    % 0.0092, 0.0468, 0.2029

    \legend{has defender, no defender}
    
    \end{axis}
\end{tikzpicture}
\caption{}%j
\end{subfigure}

\caption{\textbf{Face Identification and Object Recognition Results:} This figure shows the results for 6 configurations of the \texttt{photo-to-id} classifier network (charts a-e) and \texttt{image-to-object} classifier network (charts f-j), measured in terms of five performance criteria. The configurations includes 2 defender configurations by 3 partition point selections: $H_0$ (\texttt{pool1} activation), $H_1$ (\texttt{pool2} activation), $H_2$ (\texttt{pool3} activation). 
The 5 performance criteria include the privacy partition model inference accuracy; the reprint accuracy; the structural similarity index (SSIM) of recovered and originals images; the mean squared error (MSE) of recovered and originals images; and the deep perceptual distance (DPD) of recovered and originals images.}
\label{partition:fig:face-cifar-results}

\end{mdframed}
\end{figure*}
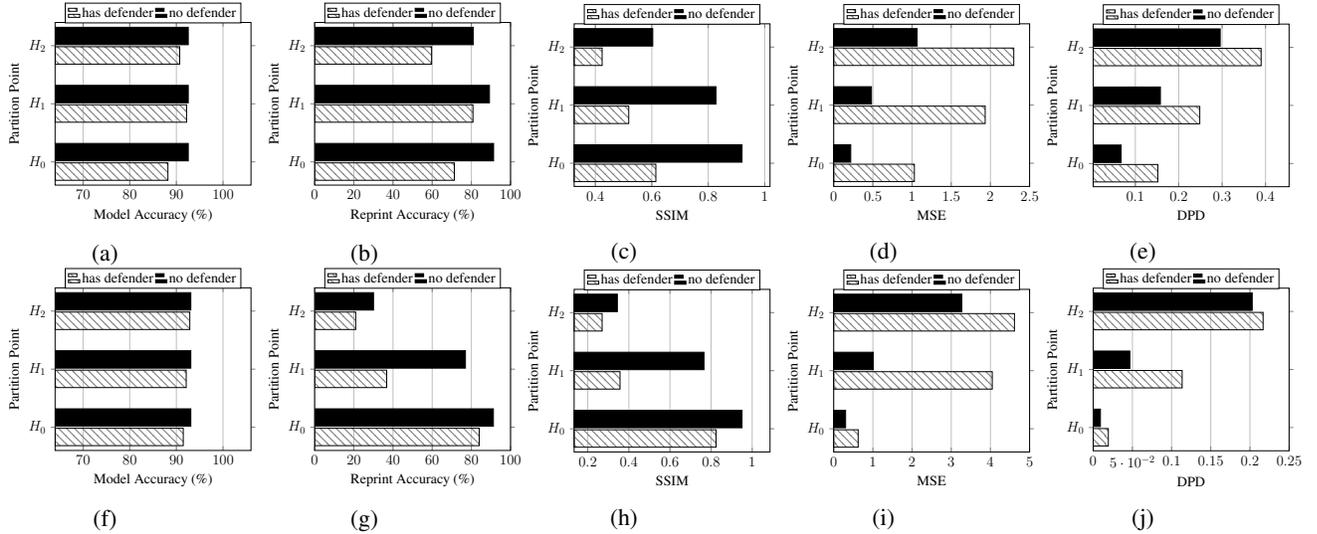

In the model inference phase, we use three types of attacker models mentioned above to recover input images. We measure the SSIM, MSE, deep perceptual distance and reprint accuracy between the input images and the recovered images by all three attackers to evaluate the quality of attacks.
We also keep track of the model inference accuracy. We conduct all our experiments
with and without the presence of the defender for comparison. \figref{partition:fig:face-cifar-results} (a)-(e) shows the results of the best attacks. 
Note that we run all of the experiments five times and compute the average.

As we can see from \figref{partition:fig:face-cifar-results} (a), the privacy partitionings still keeps the model inference accuracy at a good level: in all layers we test, the model accuracy is greater than $88\%$. We also notice that by putting the privacy partitionings in ``deeper'' layers, the degradation of model performance is less than that of putting the privacy partitioning in the first layer \texttt{pool1}.

\figref{partition:fig:face-cifar-results} (b) shows for all privacy partitionings in all layers, our framework makes it hard for the deep neural network classifier to classify them correctly (reprint accuracy). The over trend of reprint accuracy decrease as we deploy the privacy partitionings in the deeper layers. 

In \figref{partition:fig:face-cifar-results} (c)-(e), we consider three perceptual metrics discussed in \secref{partition:sec:exp-eval-metrics} to evaluate the quality of recovered images. It is obvious that the defender deteriorates the quality of recovered input images for all perceptual metrics concerned. The results also suggest that by deploying the privacy partitionings in the deeper layers, the recovered image by more indistinguishable to human. \figref{partition:fig:face-id-recover} shows the visualization results.

Overall, since our goal is to maintain the model inference accuracy while keeping recovery error of the attacker as high as possible, combined with the all five metrics we consider, deeper layers such as \texttt{pool3} are the best positions for the privacy partitioning. However, deploying more layers in the local domain would compromise the intellectual property of the deep learning models. The model providers and users should have a trade-off between model privacy and data privacy. We confirm this conclusion in \secref{partition:sec:exp-cifar10}.

 \textit{(2) Obfuscation of other private attributes}\hfill
 
 We demonstrate that our method can harden the attacker from learning other sensitive attributes from the recovered inputs. In this experiment, we perform gender estimation for the recovered images of faces by the attacker. We still use LFW dataset in this experiment since there are gender labels for each face image. Note that all the attacker models are trained for different privacy partitionings in the previous section.
 
 The standard pipelines of gender estimation are divided into two steps: The first step is to detect the face location in the images; The second step is to perform gender classification for each detected face. For the first step, we run the most commonly used HOG+SVM face detection algorithm to extract faces from images; For the second step, we use the pretrained Wide Residual Network~\cite{he2016deep} for gender classification. The results are shown in~\tabref{partition:tab:id-gender}.

 \begin{table}[ht]
\centering
\begin{footnotesize}

\ra{1.5} % manually adjust the line height of table rows

\caption{{\bf Gender Estimation Results}}
\label{partition:tab:id-gender}
\begin{tabular}{|c|l|l|c|l|l|c|l|l|}

\hline

\multicolumn{9}{|c|}{No Defense}                                                                                 \\ \hline
\multicolumn{3}{|c|}{\texttt{pool1}}          & \multicolumn{3}{c|}{\texttt{pool2}}          & \multicolumn{3}{c|}{\texttt{pool3}}          \\ \hline
\multicolumn{3}{|c|}{\#CG / \#DF / \#TI} & \multicolumn{3}{c|}{\#CG / \#DF / \#TI} & \multicolumn{3}{c|}{\#CG / \#DF / \#TI} \\ \hline
\multicolumn{3}{|c|}{414 / 459 / 474}    & \multicolumn{3}{c|}{408 / 448 / 474}    & \multicolumn{3}{c|}{283 / 309 / 474}    \\ \hline
\hline

\multicolumn{9}{|c|}{Has Defense}                                                                                \\ \hline
\multicolumn{3}{|c|}{\texttt{pool1}}          & \multicolumn{3}{c|}{\texttt{pool2}}          & \multicolumn{3}{c|}{\texttt{pool3}}          \\ \hline
\multicolumn{3}{|c|}{\#CG / \#DF / \#TI} & \multicolumn{3}{c|}{\#CG / \#DF / \#TI} & \multicolumn{3}{c|}{\#CG / \#DF / \#TI} \\ \hline
\multicolumn{3}{|c|}{359 / 390 / 474}    & \multicolumn{3}{c|}{247 / 276 / 474}    & \multicolumn{3}{c|}{66 / 76 / 474}      \\ \hline

\end{tabular}
     \\ [2ex] %\justify
     \begin{center}
    This table shows the results of gender estimation from the recovered images by the attacker. \#CG is the number of correctly classified faces for gender estimation; \#DF is the number of detected faces for all images; \#TI is the total number of images.
    \end{center}
\end{footnotesize}
\end{table}

In~\tabref{partition:tab:id-gender}, we can clearly see that the number of correctly classified faces for gender estimation (\#CG) and the number of detected faces for all images (\#DF) are significantly fewer when privacy partitioning is applied for all three partition points. Thus, we conclude that our method hardens the attacker from learning other sensitive attributes from the recovered inputs.

 \textit{(3) Insight of defender's role when training the model}\hfill
 
 \begin{figure}[!ht]
\begin{mdframed}[
    tikzsetting={align=center, draw=black, thick, align=center},
    innertopmargin=10pt,innerrightmargin=10pt,innerleftmargin=10pt,innerbottommargin=10pt, topline=false,
  rightline=false,
  leftline=false,
  bottomline=false
]
    \centering
    
    \psfrag{a}[c][c][0.9]{\makecell{no defender\\(a)}}
    \psfrag{b}[c][c][0.9]{\makecell{has defender\\(b)}}

    \includegraphics[width=3.3in]{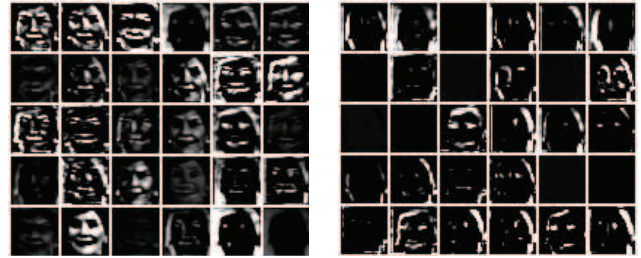}
    \caption{{\bf Hidden Layers and Defender Strategy:} This figure shows the impact of defender strategy on the hidden layer activations of the \texttt{photo-to-id} classifier network. The hidden layer activation visualizations are outputs of pooling layer $H_0$ when there is no defender present (a) and when there is a defender present (b).}
    \label{partition:fig:face-id-hidden-layers}
    
\end{mdframed}    
\end{figure}

Our experiment results demonstrate quantitatively and qualitatively the effectiveness of our method.
In order to have a more intuitive understanding of why the defender is effective, we compare the visualizations of the hidden layer activation we protect.
From the visualization comparisons, we can qualitatively check how the defender helps protect the information of hidden layer features from leaking privacy-invasive characteristics.

We design the following experiments to gain better insights of our method: given an image, we visualize the filter activations of the hidden layers where deploy the privacy partitioning in grey-scale images. We compare the ``protected'' activations with those without protection. Note that we choose the activation \texttt{pool1} and \texttt{pool2} for visualization since deeper layer features hard for a human to interpret semantically. \figref{partition:fig:face-id-hidden-layers} shows the visualization results for \texttt{pool1}. The rest of the visualization results are in \appref{partition:appendix:id}.

We can clearly see from \figref{partition:fig:face-id-hidden-layers} the differences between hidden layer activation of \texttt{pool1} with and without the defender. In \figref{partition:fig:face-id-hidden-layers} (a), we can see more clear human visually-recognizable features in more filter activation, whereas in \figref{partition:fig:face-id-hidden-layers} (b), we can only capture the features such as the basic outlines of human faces in fewer filter activations. 

From these results, we know that our method works by emphasizing ``key'' features for the original task and deactivating ``redundant'' but maybe ``sensitive'' features that do not help much for inference to increase privacy loss. Combined with the model classification accuracy results shown in~\figref{partition:fig:face-cifar-results} (a), we can conclude that performing the privacy partitioning in the proper layer not only maintain the model inference performance (emphasizing ``key'' features) but also make it harder for the attacker to ``steal'' more information from input images (deactivating ``redundant'' but  maybe ``sensitive'' features).

\subsection{Privacy Partitioning in very Deep CNN using CIFAR-10 Dataset}\label{partition:sec:exp-cifar10}
In this experiment, we continue to demonstrate the effectiveness of our framework in very deep CNN model. 
We use CIFAR-10 dataset~\cite{krizhevsky2009learning} in the experiment. CIFAR-10 is one of the benchmark datasets for object recognition and it contains $50,000$ training images and $10,000$ test images. We use VGG-19~\cite{simonyan2014very} as the model in this experiment. VGG-19 is one the most commonly used and state-of-the-art deep CNN models. The hyperparameters used in training and network architectures of the defender and the attackers can be found in \appref{partition:appendix:cifar10}. 

 \textit{(1) Deployment strategy for very deep CNN model}\hfill
 
 We conduct parallel experiments similar to \secref{partition:sec:exp-face-id} to verify the deployment strategy for very deep CNN model. The quantitative results are shown in \figref{partition:fig:face-cifar-results} (f)-(j) and the visualization results are shown in \appref{partition:appendix:cifar10}.
 
 The results in \figref{partition:fig:face-cifar-results} in (f)-(j) demonstrate that our method works very well in the very deep CNN model. For the privacy partitioning deployments in all layers, the decrease in model accuracy is less than $2\%$: $1.6\%$, $1.0\%$, $0.2\%$ drop in \texttt{pool1}, \texttt{pool2}, \texttt{pool3}, respectively (see in \figref{partition:fig:face-cifar-results} (f)). In the mean time, the indistinguishability of the recovered images by the best attacker decrease as the privacy partitioning is deployed in ``deeper'' layers (see in \figref{partition:fig:face-cifar-results} (g)-(j)). Thus, we can draw the similar conclusion about the privacy partitioning in deep CNN model as in \secref{partition:sec:exp-face-id}.
 
 \textit{(2) Continue training on the remote layers}\hfill
 
 Since our partition strategy fits well the concept fine-tuning~\cite{yosinski2014transferable} in DNNs. We can continue fine-tuning the remote layers for better model performance while keeping the local layers unchanged. We fine-tune our pretrained model in all settings we tested before. 
 We observe $0.2\%-0.3\%$ increase in model inference accuracy for all privacy partitioning deployments in \texttt{pool1}, \texttt{pool2}, \texttt{pool3}. 
 
 \textit{(3) Comparison with differential private image publication}\hfill
 
 We compare the privacy partitioning with differential private image pixelation in~\cite{fan2018image}. Their work extends standard differential privacy notion to image data. Their threat model is similar to us: they consider that the image owners wish to share one or more images to the untrusted recipients. They propose the notion of  $m$-neighborhood for image data: two images are neighboring images if they have the same dimension and they differ by at most $m$ pixels. They argue that the $m$-differences of pixels can help to protect the presence or absence of any sensitive information in the image (e.g. object, text, or person). They propose a Differentially Private Pixelation algorithm that achieves $\epsilon$-differential privacy (see in \appref{partition:appendix:dp} for details). They validate the effectiveness of the algorithm by SSIM.
 
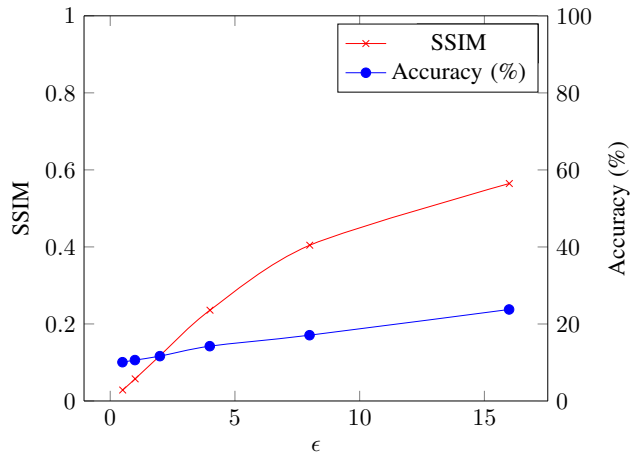
\begin{figure}[!ht]
\begin{mdframed}[
    tikzsetting={align=center, draw=black, thick, align=center},
    innertopmargin=10pt,innerrightmargin=10pt,innerleftmargin=10pt,innerbottommargin=10pt, topline=false,
  rightline=false,
  leftline=false,
  bottomline=false
]
    \centering
    \begin{tikzpicture}[scale=0.9]
    \centering
% \pgfplotsset{
%     scale only axis,
%     scaled x ticks=base 10:3,
%     xmin=0, xmax=0.06
% }
% first y-axis
\begin{axis}[
  axis y line*=left,
  ymin=0, ymax=1,
  xlabel=$\epsilon$,
  ylabel=SSIM,
]
\addplot[smooth,mark=x,red]
  coordinates{
    (0.5,0.0284)
    (1,0.0574)
    (2,0.1179)
    (4,0.2357)
    (8,0.4044)
    (16, 0.5646)
}; \label{plot_one}
% 0.5 1 2 4 8 16
% 0.0284, 0.0574, 0.1179, 0.2357, 0.4044, 0.5646
\addlegendentry{SSIM}
\end{axis}
% second y-axis
\begin{axis}[
  axis y line*=right,
  axis x line=none,
  ymin=0, ymax=100,
  ylabel= Accuracy (\%),
]
\addlegendimage{/pgfplots/refstyle=plot_one}
\addlegendentry{SSIM}
\addplot[smooth,mark=*,blue]
  coordinates{
    (0.5,10.08)
    (1,10.63)
    (2,11.65)
    (4,14.24)
    (8,17.09)
    (16,23.77)
}; \addlegendentry{Accuracy (\%)}
\end{axis}
% 0.5 1 2 4 8 16
% 10.08, 10.63, 11.65, 14.24, 17.09, 23.77

\end{tikzpicture}

\caption{{\bf Differentially Private Image Pixelation Results:} the figure shows that as privacy parameter $\epsilon$ in differentially private image pixelation become larger, both SSIM and classification accuracy increase.}
\label{partition:fig:dp}

\end{mdframed}    
\end{figure}

 We apply their algorithm for the CIFAR-10 dataset in the model inference phase on the pretrained model. We set the pixelation grid cell length to be 2 and $m$ to be 1 to minimize the negative effect image utility~\cite{fan2018image}. We change the value of $\epsilon$ to see how SSIM and the classification accuracy change accordingly. The results are shown in \figref{partition:fig:dp}. 
 
As we can see from \figref{partition:fig:dp}, the classification accuracy drops significantly. Compared to our method (see \figref{partition:fig:face-cifar-results} (f) and (g)), our method achieve similar SSIM while still keeping the classification accuracy nearly unchanged. This demonstrates our method is more practical in the real world scenario.

\section{Related Work}\label{partition:sec:related-work}
%Related Work

This line of research spans several active research areas. In this section, we discuss the related research in the areas of machine learning privacy threats and privacy-preserving solutions of deep learning with a comparison to our work.

\subsection{Machine Learning Privacy Threats}\label{Machine-Learning-Privacy-Threats}
Data privacy of machine learning has been an active research topic for long. In one of the newest works, Hitaj, \etal~\cite{HitajAP17} posit that there are fundamental limitations to the level of privacy that can be achieved using a decentralized approach to training deep learning models. This work introduces a side-channel attack to decentralized deep learning that leverages a Generative Adversarial Network (GAN) -- \ie a model that produces samples from the same distribution as the training set (prototypical samples) given model parameters -- to disclose distinctive, potentially privacy-invasive attributes of training data from other honest participants.

Song \etal~\cite{song2017machine} consider a malicious machine learning services provider who supplies model-training code to the data-holder. They demonstrate that the model capacity of deep neural networks can be abused to secretly encode the subset of the training set while still keeping the predictive power of deep learning models.

Fredrikson \etal~\cite{fredrikson2015model} explore model inversion attack: they show that model inversion could lead to unexpected privacy threats by leveraging confidence values given by machine learning models. They evaluate their attack over decision trees for ML-as-a-service and simple facial recognition
models such as MLP and Softmax Regression.

Shokri \etal~\cite{shokri2017membership} study membership inference attacks: they assume a black-box access to a machine learning inference model and determine whether a labeled data instance appears in the training data that is used to train the model. Other membership inference problems are studied in~\cite{homer2008resolving,backes2016membership,dwork2015robust,pyrgelis2017knock}

Model privacy in machine learning is another important research area. Tram{\`e}r \etal~\cite{tramer2016stealing} demonstrate the feasibility of duplicating the functionality of machine learning models such as decision tree and Logistics Regression in ML-as-a-service (MLaaS) system. They term it as \textit{model extraction attack}. Wang \etal ~\cite{wang2018stealing} propose \textit{hyperparameter stealing attacks} and demonstrate its effectiveness theoretically and empirically in machine learning models like logistics regression and support vector machine.

%\todo{add model IP}

\subsection{Privacy-preserving Solutions of Deep Learning}

\subsubsection{Differential privacy} 
Differential privacy originates from the domain of database and it provides formal privacy guarantees for each individual data record in the database. Differential privacy has been applied to ensuring the privacy of training data of deep learning to protect against the case that the model provider could learn from model parameters whether the individual data is present or not~\cite{abadi2016deep,papernot2016semi}. 

However, our threat model is different and it is hard to directly apply the standard differential privacy notion in our case: for the task of image publication for MLaaS, we often send image data to the model provider to request for the certain services for that particular image data and the information leakage by directly sending to the model provider is not guarantee by differential privacy. In summary, standard differential privacy is for protecting the privacy of individual entries in a confidential database (which is well defined for training a machine learning model), whereas the privacy partitioning is for training a deep network in such a way that inputs for classification are protected. They are complementary but not directly comparable. In summary, differential privacy provides privacy/anonymity for the individual samples training database while the privacy partitioning provides privacy during model inference.

Erlingsson \etal~\cite{erlingsson2014rappor} propose a Randomized Aggregated Privacy-Preserving Ordinal Response (RAPPOR) to provide strong privacy guarantees for crowdsourcing population statistics from end-users. However, a user might send a single data item for MLaaS system. Thus, RAPPOR cannot be applied to this scenario if a user might send a single data item for MLaaS.

Fan \etal~\cite{fan2018image} also extend the standard notion of differential privacy to image privacy for image publication. However, their method suffers from low data utility compared with our method (\secref{partition:sec:exp-cifar10}).

\subsubsection{Collaborative Training}
Shokri \etal~\cite{shokri2015privacy} propose a collaborative deep learning framework to render multiple parties to learn a deep neural network without uploading their data to the remote server in the model learning phase. Each local party has a copy of the model that can upload and download parameters during training, so that the model is trained without uploading the data to the central server party. However, the deployment of the model in the local side would increase the local computing powers and leak the model privacy. Our approach complements this framework by deploying part of the model in the local domain in the model inference phase. 

McMahan \etal~\cite{mcmahan2016communication} propose a federated averaging algorithm to protect the privacy of training data. Instead of uploading data directly to a remote server, the client trains model locally and uploads updated parameters to the central model. Still, our work complements it since we aim to protect data privacy in the model inference phase.

\subsubsection{Cryptography for Deep Learning}
Cryptography-based protocols have long been used in machine learning models to protect data privacy~\cite{bost2015machine, nikolaenko2013privacy, nikolaenko2013matrix, gilad2016cryptonets}. Liu \etal~\cite{liu2017oblivious} present cryptography-based oblivious protocols to protect data privacy in model inference phase for deep learning model. They design oblivious protocols for linear transformations, popular activation functions and pooling operations using secret sharing and garbled circuits in online prediction phase and perform request-independent operations using homomorphic encryption together with single instruction multiple data (SIMD) batch processing technique. Since the method requires no change in the pretrained model, our approach is complementary with theirs. Furthermore, our method is compatible with all cryptography-based protocols in principle.

\section{Conclusion}\label{partition:sec:conclusion}
%Conclusion}\label{sec:conclusion}

We evaluate the effectiveness of the proposed privacy partitioning framework and find experimentally that it a promising method for significantly reducing the capacity for an adversary with access to intermediate layer activation or a significant portion of a deep network topology to conduct input recovery attacks. We develop and evaluate the privacy partitioning framework in the context of a novel threat model and often used deployment context. 
Future work in this research will explore integration with compatible deep learning privacy protections, integration with software and hardware security modules used to secure a local domain, as well as the formal guarantee of our method in different neural network configurations and data distributions.

%\section*{Acknowledgments}
%\import{sections/}{00-acknowledgment}



%\bibliographystyle{IEEEtranS}

%\newpage
\bibliography{library}
\bibliographystyle{IEEEtran}

\appendix
\subsection{Hyperparamter setting in MNIST experiment}\label{partition:appendix:mnist}
In this section, we will describe some of the experiment details for MNIST dataset. Here are all attacker networks used in our MNIST experiment:
\begin{itemize}
\item Attacker $\circled{1}$: \texttt{800}  $\rightarrow$ \texttt{Relu} $\rightarrow$ \texttt{800} $\rightarrow$ \texttt{Sigmoid}

\item Attacker $\circled{2}$: \texttt{800}  $\rightarrow$ \texttt{Relu} $\rightarrow$ \texttt{dropout(0.1)} $\rightarrow$ \texttt{800} $\rightarrow$ \texttt{Sigmoid}
\item Attacker $\circled{3}$: \texttt{800}  $\rightarrow$ \texttt{Tanh} $\rightarrow$ \texttt{800} $\rightarrow$ \texttt{Sigmoid}
\item Attacker $\circled{4}$:  \texttt{800}  $\rightarrow$ \texttt{Sigmoid} $\rightarrow$ \texttt{800} $\rightarrow$ \texttt{Sigmoid}
\item Attacker $\circled{5}$: \texttt{512}  $\rightarrow$ \texttt{Relu} $\rightarrow$ \texttt{512} $\rightarrow$ \texttt{Sigmoid}
\item Attacker $\circled{6}$: \texttt{1024}  $\rightarrow$ \texttt{Relu} $\rightarrow$ \texttt{1024} $\rightarrow$ \texttt{Sigmoid}
\item Attacker $\circled{7}$: \texttt{1-D conv}  $\rightarrow$ \texttt{Relu} $\rightarrow$ \texttt{800} $\rightarrow$ \texttt{Sigmoid}
\item Attacker $\circled{8}$: \texttt{784} $\rightarrow$ \texttt{Sigmoid}
\end{itemize}  

Here are the defender models used in the multiple defender training experiments: 

\begin{itemize}
\item Defender $\circled{1}$: \texttt{800}  $\rightarrow$ \texttt{Tanh} $\rightarrow$ \texttt{800} $\rightarrow$ \texttt{Sigmoid}
\item Defender $\circled{2}$: \texttt{800}  $\rightarrow$ \texttt{Sigmoid} $\rightarrow$ \texttt{800} $\rightarrow$ \texttt{Sigmoid}
\item Defender $\circled{3}$: \texttt{1-D conv}  $\rightarrow$ \texttt{Relu} $\rightarrow$ \texttt{800} $\rightarrow$ \texttt{Sigmoid}
\item Defender $\circled{4}$: \texttt{784} $\rightarrow$ \texttt{Sigmoid}
\end{itemize}

\subsection{Hyperparamter setting in face ID experiment}\label{partition:appendix:id}

We use the CNN model for Face ID recogition. The model used for face ID experiment is:

\texttt{conv2d 5$\times$5} $\rightarrow$ \texttt{conv2d 5$\times$5} $\rightarrow$ \texttt{maxpool 2$\times$2 (pool1)}  $\rightarrow$ \texttt{conv2d 3$\times$3} $\rightarrow$ \texttt{conv2d 3$\times$3} $\rightarrow$ \texttt{maxpool 2$\times$2 (pool1)} $\rightarrow$ \texttt{conv2d 3$\times$3} $\rightarrow$ \texttt{conv2d 3$\times$3} $\rightarrow$ \texttt{maxpool 2$\times$2 (pool3)} $\rightarrow$ \texttt{conv2d 3$\times$3} $\rightarrow$ \texttt{conv2d 3$\times$3} $\rightarrow$ \texttt{maxpool 2$\times$2} $\rightarrow$ \texttt{512} $\rightarrow$ \texttt{dropout(0.5)} $\rightarrow$ \texttt{512} $\rightarrow$ \texttt{dropout(0.5)} $\rightarrow$ \texttt{512}

Note that each convolutional layer is followed by the Relu activation and batch-normalization layer.

The design of defender model depends on which layer for partition (\eg in our experiment, we choose the the outputs of first three pooling layers \texttt{pool1}, \texttt{pool2}, and \texttt{pool3} for partition). For example, if we choose \texttt{pool2} for partition, the architecture of defender network would be the reversed version of local layers:  

\texttt{deconv2d 3$\times$3 (stride 2)} $\rightarrow$ \texttt{conv2d 3$\times$3} $\rightarrow$ \texttt{Relu} $\rightarrow$ \texttt{deconv2d 5$\times$5 (stride 2)} $\rightarrow$ \texttt{deconv2d 5$\times$5} $\rightarrow$ \texttt{tanh} 

Note that this type of architecture resembles the design strategy of an convolutional autoencoder~\cite{hinton2006reducing, erhan2010does}. The attacker networks also depend on which layer for partition since the input dimension might be different from layer to layer. For example, if we choose \texttt{pool2} for partition, the attacker is

\begin{itemize}
    \item Attacker $\circled{1}$ (\texttt{DECONV ATTACKER}): \texttt{deconv2d 3$\times$3 (stride 2)} $\rightarrow$ \texttt{conv2d 3$\times$3} $\rightarrow$ \texttt{Relu} $\rightarrow$ \texttt{deconv2d 5$\times$5 (stride 2)} $\rightarrow$ \texttt{deconv2d 5$\times$5} $\rightarrow$ \texttt{tanh}
    \item Attacker $\circled{2}$ (\texttt{FC ATTACKER}): \texttt{4096} $\rightarrow$ \texttt{Relu} $\rightarrow$ \texttt{12288} $\rightarrow$ \texttt{Tanh} 
    \item Attacker $\circled{3}$ (\texttt{SPARSE FC ATTACKER}): \texttt{4096} $\rightarrow$ \texttt{Relu} $\rightarrow$ \texttt{dropout(0.5)} $\rightarrow$ \texttt{12288} $\rightarrow$ \texttt{Tanh}
\end{itemize}

Note that these types of attackers  are chosen since they are the most commonly used decoder architecture in auto-encoder designs~\cite{hinton2006reducing} and they cover the most commonly used operations in deep neural networks in the area of computer vision.

\begin{figure}[!ht]
\begin{mdframed}[
    tikzsetting={align=center, draw=black, thick, align=center},
    innertopmargin=10pt,innerrightmargin=10pt,innerleftmargin=10pt,innerbottommargin=10pt, topline=false,
  rightline=false,
  leftline=false,
  bottomline=false
]
    \centering
    
    \psfrag{a}[c][c][0.9]{\makecell{no defender\\(a)}}
    \psfrag{b}[c][c][0.9]{\makecell{has defender\\(b)}}

    \includegraphics[width=3.3in]{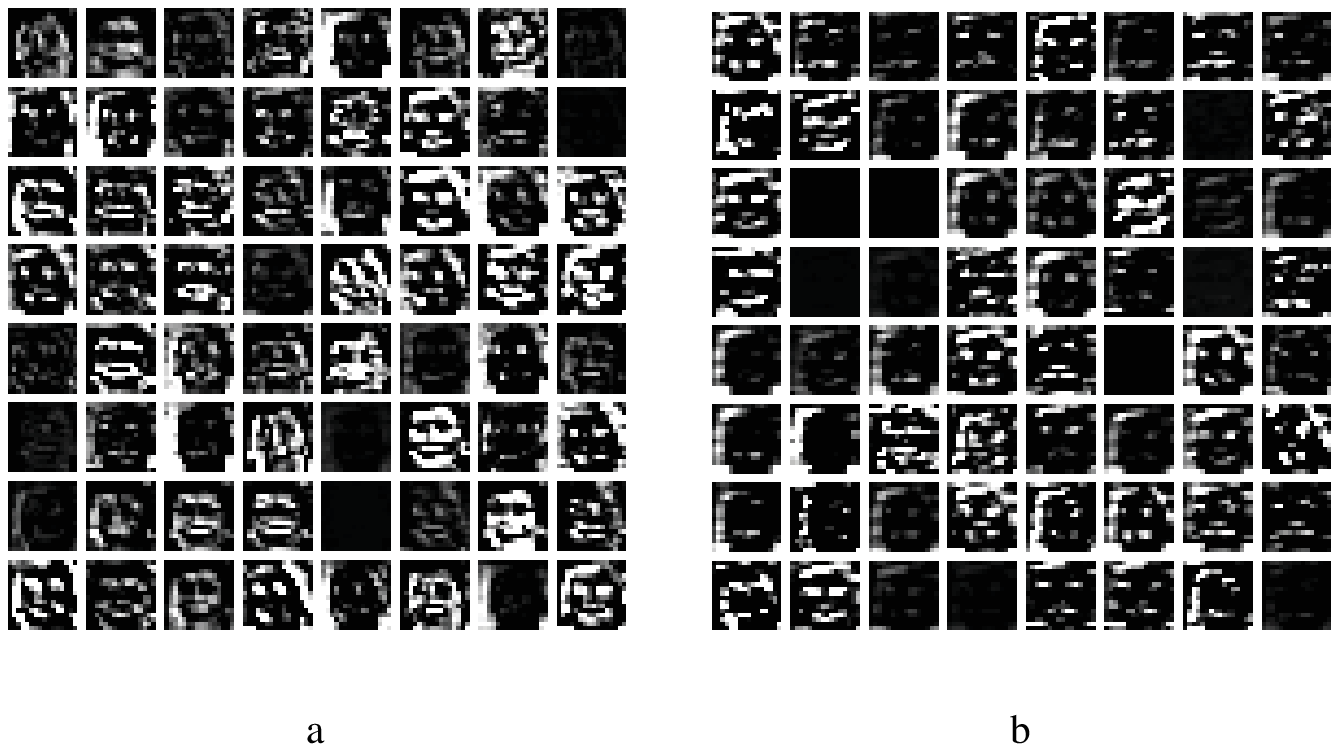}
    \caption{{\bf Hidden Layers and Defender Strategy II:} This figure shows the impact of the defender strategy on the \texttt{pool2} activations (hidden layer $H_1$) of the \texttt{photo-to-id} classifier network. Applying the defender strategy (b) results in more information loss to the intermediate layer features a compared to training normally (a).}
    \label{partition:fig:face-id-hidden-layers-pool2}
\end{mdframed}    
\end{figure}

\subsection{Hyperparamter setting in CIFAR10 experiment}\label{partition:appendix:cifar10}
We used batch-normalized VGG-19~\cite{simonyan2014very} for the classification model. We train the model with random crop and random horizontal flip data augmentation techniques for 150 epochs. The initial learning rate is 0.01 with a decay factor of 0.1 for every 50 epochs. we use SGD algorithm with momentum 0.9 and $l$-2 penalty weight decay 0.0005.

The design strategies of defender and attacker networks are similar in the face ID experiment. For example, if we choose the second pooling layer \texttt{pool2} for partition, the defender network is:

\texttt{deconv2d 3$\times$3 (stride 2)} $\rightarrow$ \texttt{BatchNorm}  $\rightarrow$ \texttt{Relu} $\rightarrow$ \texttt{deconv2d 3$\times$3 (stride 2)} $\rightarrow$ \texttt{BatchNorm} $\rightarrow$ \texttt{Relu} $\rightarrow$ \texttt{deconv2d 3$\times$3} $\rightarrow$ \texttt{BatchNorm}

And the attacker networks are:

\begin{itemize}
    \item Attacker $\circled{1}$ (\texttt{DECONV ATTACKER}): \texttt{deconv2d 3$\times$3 (stride 2)} $\rightarrow$ \texttt{BatchNorm}  $\rightarrow$ \texttt{Relu} $\rightarrow$ \texttt{deconv2d 3$\times$3 (stride 2)} $\rightarrow$ \texttt{BatchNorm} $\rightarrow$ \texttt{Relu} $\rightarrow$ \texttt{deconv2d 3$\times$3} $\rightarrow$ \texttt{BatchNorm} 
 
 \item Attacker $\circled{2}$ (\texttt{FC ATTACKER}): \texttt{1024} $\rightarrow$ \texttt{Relu} $\rightarrow$ \texttt{3072} 
    \item Attacker $\circled{3}$ (\texttt{SPARSE FC ATTACKER}): \texttt{1024} $\rightarrow$ \texttt{Relu} $\rightarrow$ \texttt{dropout(0.5)} $\rightarrow$ \texttt{3072}
\end{itemize}

For the outputs of the attacker networks, we will also clip them to match the normalized image pixel range. 
\begin{figure*}[!ht]
\begin{mdframed}[
    tikzsetting={align=center, draw=black, thick, align=center},
    innertopmargin=10pt,innerrightmargin=10pt,innerleftmargin=10pt,innerbottommargin=10pt, topline=false,
  rightline=false,
  leftline=false,
  bottomline=false
]
    \centering
    
    % defender configurations
    \psfrag{a}[c][c][0.9]{\makecell{no defender\\(a)}}
    \psfrag{b}[c][c][0.9]{\makecell{has defender\\(b)}}
    \psfrag{c}[c][c][0.9]{\makecell{Airplane}}
    
    \psfrag{d}[c][c][0.9]{\makecell{no defender\\(c)}}
    \psfrag{e}[c][c][0.9]{\makecell{has defender\\(d)}}
    \psfrag{f}[c][c][0.9]{\makecell{Bird}}
    
    \psfrag{g}[c][c][0.9]{\makecell{no defender\\(e)}}
    \psfrag{h}[c][c][0.9]{\makecell{has defender\\(f)}}
    \psfrag{i}[c][c][0.9]{\makecell{Horse}}
    
    \psfrag{j}[c][c][0.9]{\makecell{no defender\\(g)}}
    \psfrag{k}[c][c][0.9]{\makecell{has defender\\(h)}}
    \psfrag{l}[c][c][0.9]{\makecell{Ship}}

    \includegraphics[width=6.2in]{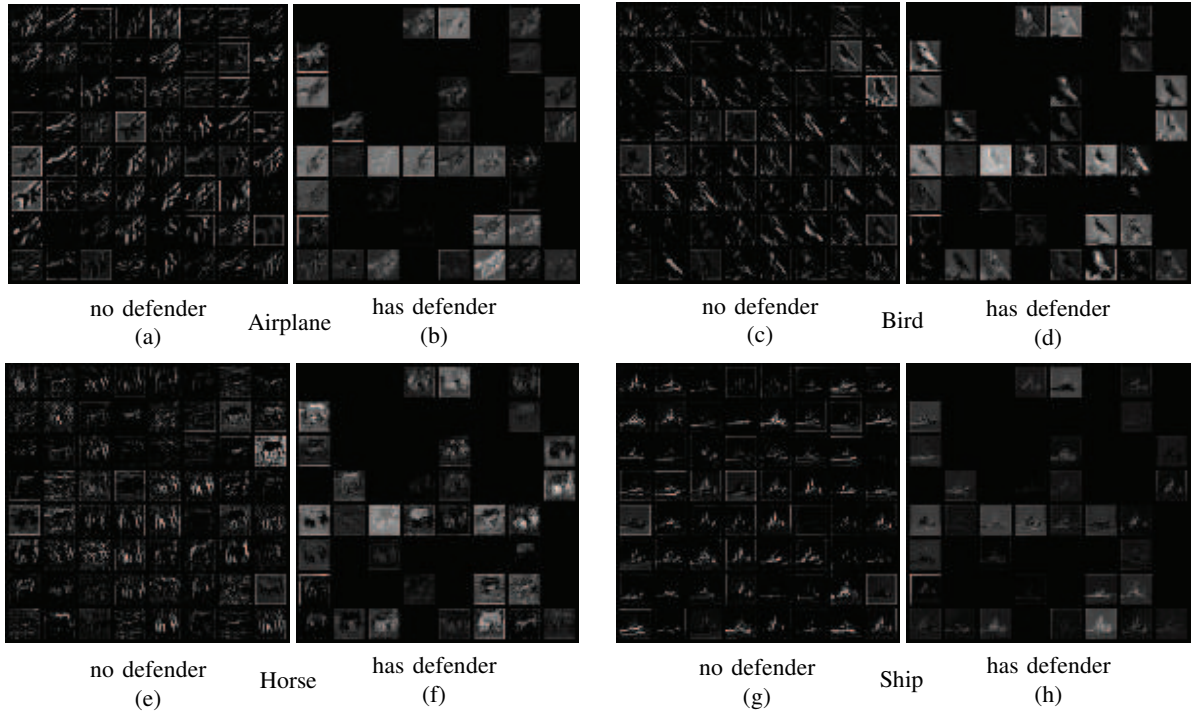}
    \caption{{\bf Hidden Layer Features of CIFAR-10 Data:} This figure shows the impact of the defender strategy on different object classes. Privacy partitioning introduces sparsity to hidden layer features to increase information loss as compared to training the deep network normally.}
    \label{cifar10-hidden}
    
\end{mdframed}    
\end{figure*}

\begin{figure*}[!ht]
\begin{mdframed}[
    tikzsetting={align=center, draw=black, thick, align=center},
    innertopmargin=10pt,innerrightmargin=10pt,innerleftmargin=10pt,innerbottommargin=10pt, topline=false,
  rightline=false,
  leftline=false,
  bottomline=false
]
    \centering
    
    % defender configurations
    \psfrag{a}[c][c][0.9]{\makecell{has\\defender}}
    \psfrag{b}[c][c][0.9]{\makecell{no\\defender}}
    
    % layer configurations
    \psfrag{c}[c][c][0.9]{(\texttt{pool1})}
    \psfrag{d}[c][c][0.9]{(\texttt{pool2})}
    \psfrag{e}[c][c][0.9]{(\texttt{pool3})}
    
    % axis labels
    \psfrag{f}[c][c][0.9]{\rotatebox{90}{defenders}}
    \psfrag{g}[c][c][0.9]{local layers}
    
    % original images
    \psfrag{h}[c][c][0.9]{original images}

    \includegraphics[width=6.2in]{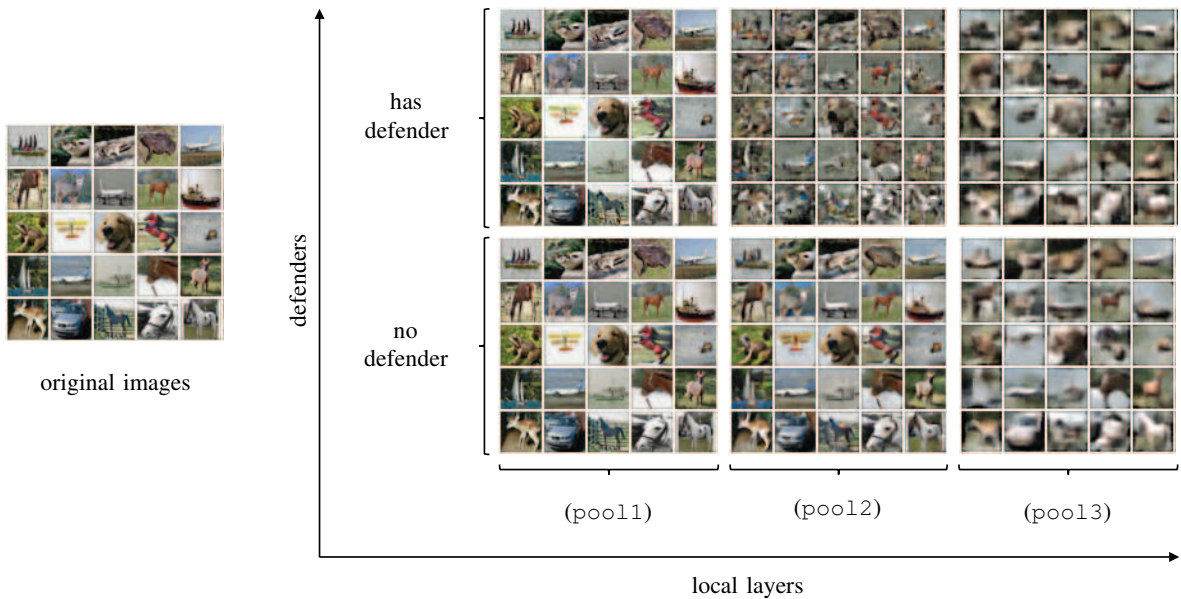}
    \caption{\textbf{CIFAR-10 Data Recovery:} This figure shows the sample images recovery visualization results. The application of the privacy partitioning hardens the input recovery for all layer configurations and the recovery error increases as more layers are included in the local partition $\Theta_l$.}
    \label{partition:fig:cifar10-recover}
    
\end{mdframed}    
\end{figure*}

\subsection{Differentially Private Image Pixelation}\label{partition:appendix:dp}

In this section we will discuss the the notion of differentially private image pixelation~\cite{fan2018image} and how to achieve it. To apply the notion of differential privacy to image, they define the notion of ``neighboring images''.

\textbf{Definition} [$m$-Neighborhood] Two images $I_1$ and $I_2$ are neighboring images if they have the same dimension and they differ by at most $m$ pixels.

To achieve $\epsilon$-differential privacy, they also propose  Differentially Private Pixelation algorithm. The details of the algorithm are in~\algref{partition:alg:dp-pixel}.

\begin{algorithm}
    \SetKwInOut{Input}{Input}
    \SetKwInOut{Output}{Output}
    \SetKwInOut{Initialize}{Initialize}
    
    \Input{Input Image $I$ with size $M \times N$, neighbouring parameter $m$, pixelation grid size $b$, privacy parameter $\epsilon$}
    \Output{Differentially private image $\tilde{I}$}
    
    Divide $I$ into $K=\lceil\frac{M}{b}\rceil  \lceil\frac{N}{b}\rceil$ cells $c_k$, where $\quad k = 1,\dots, \lceil\frac{M}{b}\rceil  \lceil\frac{N}{b}\rceil$ \;
    
    Pixelate the image 
    \begin{equation}
        \nonumber
        p(I;b) = \{\frac{1}{b^2} \sum_{(x,y) \in c_1} I(x,y),\dots, \frac{1}{b^2} \sum_{(x,y)\in c_{K}} I(x,y)\}
    \end{equation}
 
    Sample Laplacian noises $\tilde{\mathbf{N}} = \{\tilde{N}_1, \tilde{N}_2, \dots, \tilde{N}_K\}$ with means 0 and and scales $\frac{255m}{b^2\epsilon}$\; 
    
    Add noises to the pixelated image $\tilde{I} = I + \tilde{\mathbf{N}}$\;
    
    \Return $\tilde{I}$\;
    
    \caption{Differentially Private Image Pixelation \texttt{pix}}
    \label{partition:alg:dp-pixel}
\end{algorithm}

\textbf{Theorem 1} \algref{partition:alg:dp-pixel} is $\epsilon$-differential private.

\textit{Proof}. The global sensitivity of~\algref{partition:alg:dp-pixel} is $\Delta = \max_{I_1, I_2}\vert \texttt{pix}(I_1) - \texttt{pix}(I_2) \vert = \frac{255m}{b^2}$ for any neighboring images $I_1, I_2$. By \textbf{Theorem 2} in~\cite{dwork2006calibrating}, we can conclude that ~\algref{partition:alg:dp-pixel} is $\epsilon$-differential private.

\end{document}